\documentclass[aps,prb, reprint,amsmath,amssymb,longbibliography,superscriptaddress]{revtex4-2}

\usepackage{graphicx}
\usepackage{dcolumn}
\usepackage{bm}
\usepackage{multirow}
\usepackage{color}
\usepackage{amsmath}
\usepackage{gensymb}
\usepackage[colorlinks=true,citecolor=blue,linkcolor=blue, urlcolor=cyan]{hyperref} 

\usepackage{textcomp}
\usepackage{placeins}
\usepackage{braket}
\usepackage[normalem]{ulem}

\usepackage[dvipsnames]{xcolor}
\usepackage{amsmath}
\usepackage{amssymb}
\usepackage{bm}
\usepackage{epsfig}
\usepackage{graphicx}
\usepackage{color}
\usepackage{hyperref}
\usepackage{orcidlink}

\def\be{\begin{equation}}
\def\ee{\end{equation}}
\def\bea{\begin{eqnarray}}
\def\eea{\end{eqnarray}}

\begin{document}

\title{ Excitons and trions in CrSBr bilayers}
	
	\author{M. A. Semina\orcidlink{0000-0003-3796-2329}}\email{semina@mail.ioffe.ru}
	\affiliation{Ioffe Institute, 194021, St.-Petersburg, Russia}
	\author{F. Tabataba-Vakili\orcidlink{0000-0001-5911-7594}}
	\affiliation{Institute of Condensed Matter Physics, Technische Universit\"at Braunschweig, 38106 Braunschweig, Germany}
		\affiliation{Fakult\"at f\"ur Physik, Munich Quantum Center, and Center for NanoScience (CeNS), Ludwig-Maximilians-Universit\"at M\"unchen, 
80539 M\"unchen, Germany}
\affiliation{Munich Center for Quantum Science and Technology (MCQST), 80799 M\"unchen, Germany}
	\author{A. Rupp}
			\affiliation{Fakult\"at f\"ur Physik, Munich Quantum Center, and Center for NanoScience (CeNS), Ludwig-Maximilians-Universit\"at M\"unchen, 
80539 M\"unchen, Germany}
	\author{A.S. Baimuratov\orcidlink{0000-0001-8708-2405}}
			\affiliation{Fakult\"at f\"ur Physik, Munich Quantum Center, and Center for NanoScience (CeNS), Ludwig-Maximilians-Universit\"at M\"unchen, 
80539 M\"unchen, Germany}
			\affiliation{Center for Engineering Physics, Skolkovo Institute of Science and Technology, Bolshoy Boulevard 30, building 1, Moscow 121205, Russia}
	\author{A. H\"ogele\orcidlink{0000-0002-0178-9117}}
	\affiliation{Fakult\"at f\"ur Physik, Munich Quantum Center, and Center for NanoScience (CeNS), Ludwig-Maximilians-Universit\"at M\"unchen, 
80539 M\"unchen, Germany}
\affiliation{Munich Center for Quantum Science and Technology (MCQST), 80799 M\"unchen, Germany}
		\author{M.M. Glazov\orcidlink{0000-0003-4462-0749}}\email{glazov@coherent.ioffe.ru}
	\affiliation{Ioffe Institute, 194021, St.-Petersburg, Russia}
	
	\begin{abstract}
We study theoretically the neutral and charged excitons in two-dimensional semiconductors with anisotropic dispersion of charge carriers. Such a situation is realized in CrSBr-based van der Waals heterostructures. We calculate the binding energies of excitons and trions and explore their dependence on the mass ratio, dielectric screening, and interlayer distance in bilayer structures. {We also address the effects of exciton-light coupling, including the radiative decay and long-range electron-hole exchange interaction, and briefly analyze correlations between the excitons and the Fermi sea of resident electrons}. The estimates for CrSBr bilayers are in reasonable agreement with recent experiments. 
	\end{abstract}	
	\maketitle
\date{\today}

\section{Introduction}\label{sec:intro}

The family of atomically thin materials and van der Waals heterostructures has been recently widened by magnetic materials. A special interest is attracted by CrSBr that combines semiconducting and magnetic properties~\cite{Telford:2020aa,Wilson:2021aa,Klein:2023aa,Dirnberger:2023aa,Klein2024}. It is a direct gap semiconductor with a band gap of about $1.5$ --- $2.1$~eV \cite{Telford:2020aa,smolenski2024largeexcitonbindingenergy, Watson2024}, ferromagnetic intralayer order and antiferromagnetic order between the layers~\cite{Goser:1990aa,Scheie:2022aa,Lee:2021aa,Bo_2023}. It demonstrates fascinating excitonic and exciton-polaritonic effects~\cite{Dirnberger:2023aa} and strong exciton-magnon interaction~\cite{Bae:2022aa,Diederich:2023aa}. CrSBr provides a playground for studying interactions between quasiparticles of different nature: excitons, the electronic excitations, and magnons, the spin waves. Moreover, optical response of this material mediated by excitonic species provides a deep insight in magnetic arrangement and dynamics, particularly, in the presence of resident charge carriers~\cite{D3TC01216F,Tabataba-Vakili:2024aa}.

The detailed studies of intertwined systems of excitons and magnons and possible applications require a solid understanding of the basic quasiparticles underlying optical response. These are neutral excitons, the Coulomb-bound electron-hole pairs, and trions, three particle complexes formed from two electrons and a hole or two holes and an electron that appear as a result of the exciton formation in doped structure and subsequent binding with resident charge carriers~\cite{ivchenko05a,Klingshirn2012,Yu30122014,RevModPhys.90.021001,Durnev_2018,Semina_2022}. Specifics of CrSBr in this regard are: (i) weak interlayer coupling, (ii) strong in-plane anisotropy of electronic bands, and (iii) spin polarization of the band structure resulting in the spin-layer locking~\cite{PhysRevB.107.235107,Klein:2023aa,Tabataba-Vakili:2024aa}. It calls for theoretical description of excitons and trions in this material and the analysis of their energy spectrum fine structure resulting from the exchange interactions and also their coupling with the Fermi sea of resident charge carriers. The latter gives rise to the Fermi-polarons also known as Suris tetrons, effectively four-particle complexes formed by the photoelectron and photohole and an electron-hole pair in the Fermi sea~\cite{PSSB:PSSB343,PhysRevLett.112.147402,Sidler:2016aa}. As compared to the previous works devoted to the electron-hole complexes in anisotropic two-dimensional materials such as, e.g., black phosphorous,~\cite{PhysRevB.93.115314,PhysRevB.98.235401}, the specifics of CrSBr are related to the full spin polarization of bands and the presence of interlayer complexes where different charge carriers occupy different layers of the material.

In this work we develop a theory of excitons, trions and Fermi-polarons/Suris tetrons in CrSBr mono- and bilayers. The bilayer is a minimum model system that already demonstrates all the specifics of few- and multilayer CrSBr, whereas  the monolayer is a natural building block. Based on the combination of numerical diagonalization of a few-patricle Hamilonian, variational principle and analytical calculations we uncover the anisotropy effects and study the role of the spin-layer locking and broken time-reversal symmetry. We also analyze the effects of interlayer distance and effective mass anisotropy, as well as anisotropic dielectric environment, on the Coulomb-correlated complexes aiming to explore a range of possibilities in similar systems.

The paper is organized as follows: after a brief introduction (Sec.~\ref{sec:intro}) we formulate the studied model in Sec.~\ref{sec:model}. The results for the exciton and trion binding energies are presented and discussed in Sec.~\ref{sec:binding}. Section~\ref{sec:exch} briefly addresses the effects of the electron-hole exchange interaction in CrSBr bilayers.  The summary and concluding remarks are presented in Sec.~\ref{sec:concl}. {Appendix~\ref{AA} addresses the applicability of the one-dimensional approximation to describing the Coulomb complexes with anisotropic effective masses and the correlations between the excitons, trions, and Fermi sea of resident charge carriers are briefly analyzed in Appendix~\ref{sec:FP}.}

\section{Model}\label{sec:model}

\begin{figure}[b]
\includegraphics[width=\linewidth]{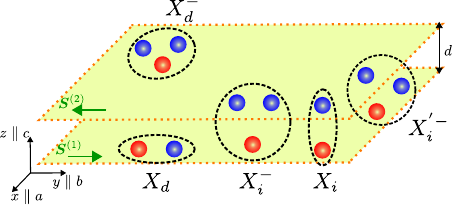}
\caption{Schematic illustration of the studied bilayer system. Green arrows $\bm S^{(1)}$ and $\bm S^{(2)}$ denote the magnetization of layers. Red and blue spheres show holes and electrons, respectively. Coulomb complexes under study are denoted as $X_d$ (direct or intralayer exciton), $X_i$ (indirect or interlayer exciton), $X_d^-$ (direct or intralayer trion), $X_i^-$ (indirect trion with two electrons in the same layer) and $X_i^{'-}$ (indirect trions with two electrons in different layers). }\label{fig:scheme}
\end{figure}

Following Refs.~\cite{Klein:2023aa,Tabataba-Vakili:2024aa} we introduce the  the set of axes with $z\parallel c$ (normal to the bilayer or to the monolayer, magnetic hard axis), $y\parallel b$ (magnetic easy axis), and
$x\parallel a$ (magnetic intermediate axis), Fig.~\ref{fig:scheme}. The intralayer ferromagnetic spin-spin interactions result in the complete spin
polarization of the Bloch states with the spins aligned along the $b$-axis in the monolayer. In a bilayer structure the spins of the first and second layer are aligned antiferromagnetically along the positive and negative directions of the $b$-axis.

In what follows we focus on a minimum $\bm k\cdot \bm p$-model description of the system and the effective mass approach. We start with the monolayer. For $D_{2h}$ symmetry of the studied structure the orbital Bloch function of the topmost valence band $v$ at each layer transforms at the $\Gamma$-point of the Brillouin zone according to the $B_{3g}$ (or $\Gamma_4^+$) irreducible representation, i.e., as the product $yz$ of the coordinates. The optically active conduction band $c_2$ corresponds, at the $\Gamma$-point, to the $B_{1u}$ ($\Gamma_3^-$) irreducible representation, its wave function transforms as the $z$ coordinate. Consequently, the optical transition $v\leftrightarrow c_2$ is allowed in $y$ polarization. Also, there is a nearby conduction band $c_1$ of the $A_g$ ($\Gamma_1^+$) symmetry and the optical transition  $v\leftrightarrow c_1$ is forbidden in the electric-dipole approximation. The spin polarization of a magnetic subsystem implies spin polarization of the electronic Bloch bands. For bi- and multilayer systems, the interlayer tunnelling  (in the absence of external magnetic field) for charge carriers is forbidden because of the opposite orientations of the spins in the layers. Thus, in a bilayer system the band structure is formed by two copies of the monolayer band structure.

\begin{figure}[b]
\includegraphics[width=\linewidth]{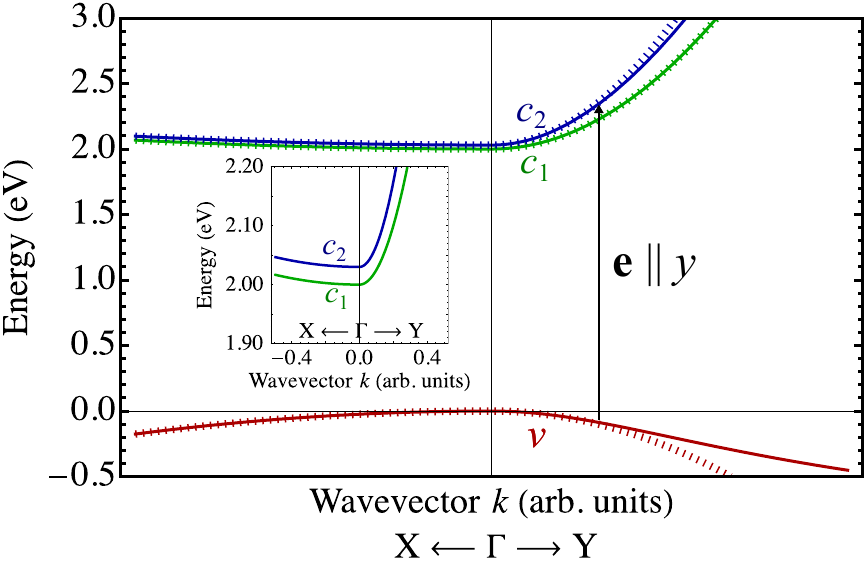}
\caption{Energy spectrum of CrSBr monolayer calculated using the three-band $\bm k\cdot\bm p$ model (solid lines) and within the effective mass approach with the masses found from Eqs.~\eqref{masses} (dotted lines). Vertical arrow denotes the allowed transition from $v$ to $c_2$ band in $y$-polarization. Inset shows a zoom-in of the conduction band dispersion in the vicinity of the $\Gamma$-point. The parameters used in the calculation roughly correspond to Ref.~\cite{Klein:2023aa} and are as follows: $E_{c1} - E_{v} = 2$~eV, $E_{c2} -E_{c1} = 30$~meV, $\hbar p_{cv}/m_0 =-1.5$~eV, $\tilde m_{1,2;x} = -1.157m_0$, $\tilde m_{1,2;y}=0.25m_0$, $\tilde m_{v,x}=-0.74m_0$, $\tilde m_{v,y} = -m_0$.}\label{fig:KP}
\end{figure}

Correspondingly, the only non-zero interband momentum matrix element at the $\Gamma$-point of the Brillouin zone is
\begin{equation}
\label{pcv}
p_{cv} = \langle B_{1u} |\hat{p}_y |B_{3g}\rangle.
\end{equation}
It allows us to construct a minimum $\bm k\cdot \bm p$ model. The three-band $\bm k\cdot\bm p$ Hamiltonian for a monolayer reads
\begin{equation}
\label{kp}
\mathcal H = \begin{pmatrix}
E_{c_2}(\bm k) & 0 & \frac{\hbar}{m_0} k_y p_{cv}\\
0& E_{c_1}(\bm k) & 0 \\
\frac{\hbar}{m_0} k_y p_{cv}^* & 0 & E_v(\bm k)
\end{pmatrix}.
\end{equation}
Here $\bm k$ is the wave vector of the electron in the CrSBr $(xy)$ plane,
\begin{subequations}
\label{E:diag}
\begin{align}
E_{c_2}(\bm k) = E_{c2}+ \frac{\hbar^2 k^2}{2m_0} + \frac{\hbar^2 k_x^2}{2\tilde m_{2,x}} + \frac{\hbar^2 k_y^2}{2\tilde m_{2,y}},\\
E_{c_1}(\bm k) = E_{c1} + \frac{\hbar^2 k^2}{2m_0} + \frac{\hbar^2 k_x^2}{2\tilde m_{1,x}} + \frac{\hbar^2 k_y^2}{2 \tilde m_{1,y}},\\
E_v(\bm k) = E_{v} + \frac{\hbar^2 k^2}{2m_0} + \frac{\hbar^2 k_x^2}{2\tilde m_{v,x}} + \frac{\hbar^2 k_y^2}{2 \tilde m_{v,y}},
\end{align}
\end{subequations}
$E_{v}$, $E_{c1}$, and $E_{c2}$ are the energies of the corresponding band edges, $m_0$ is the free electron mass, $\tilde m_{i,\alpha}$ ($i=1,2,v$, $\alpha=x$ or $y$) are the remote band contributions to the corresponding effective masses. Hereafter we mainly focus on the excitons and trions formed from the valence band holes and the  electrons in the conduction band $c_2$. Within such a model the effective masses of the electrons and holes are highly anisotropic:
\begin{subequations}
\label{masses}
\begin{align}
\frac{1}{m_x^e} = \frac{1}{m_0} + \frac{1}{\tilde m_{2,x}},\\
\frac{1}{m_x^h} = -\frac{1}{m_0} - \frac{1}{\tilde m_{v,x}},\\
\frac{1}{m_y^e} = \frac{1}{m_0} + \frac{1}{\tilde m_{2,y}} + \frac{2 |p_{cv}|^2}{m_0^2(E_{c_2} - E_v)},\\
\frac{1}{m_y^h} = -\frac{1}{m_0} - \frac{1}{\tilde m_{v,y}} + \frac{2 |p_{cv}|^2}{m_0^2(E_{c_2} - E_v)}.
\end{align}
\end{subequations} 
One can see that for the propagation along the $x$-axis the effective masses are determined by the free-electron mass and remote band contributions. By contrast, the effective masses along the $y$-axis are much lighter because of the $\bm k\cdot\bm p$ mixing of the valence and conduction bands. Indeed, as atomistic calculations show~\cite{Klein:2023aa}, the effective masses are $m_x^e=7.31 m_0$, $m_y^e=0.14 m_0$, $m_x^h=2.84 m_0$, $m_y^h=0.45 m_0$. Consequently, for excitons, the reduced effective masses [$\mu_{\alpha} = m_\alpha^e m_\alpha^h/(m_\alpha^e + m_\alpha^h)$, $\alpha=x$ or $y$] are $\mu_x\approx 2 m_0$, $\mu_y\approx 0.1 m_0$ leading also to  high anisotropy of the exciton wave function. In Ref.~\cite{smolenski2024largeexcitonbindingenergy} the values $m_x^e=12.26 m_0$, $m_y^e=0.48 m_0$, $m_x^h=3.75 m_0$, $m_y^h=0.17 m_0$ leading to $\mu_x\approx 2.9 m_0$, $\mu_y\approx 0.12 m_0$ were measured experimentally leading to a similar value of the exciton reduced mass anisotropy.  The energy dispersion calculated from the diagonalization of the $\bm k\cdot \bm p$ Hamiltonian~\eqref{kp} and using the parabolic (effective mass) approximation with the masses given by Eqs.~\eqref{masses} is shown in Fig.~\ref{fig:KP}. One can see a pronounced anisotropy of the energy spectrum and good agreement between the effective mass and $\bm k\cdot \bm p$ models. Hence, we use the effective mass approach in what follows.

Within the effective mass approach the excitons are described by the Schr\"odinger equation for the relative motion smooth envelope $\psi_{X}(\bm r)$ with the Hamiltonian
\begin{equation}
\label{H:x}
\mathcal H_{X} = -\sum_{\alpha=x,y} \frac{\hbar^2}{2\mu_\alpha} \frac{\partial^2}{\partial r_\alpha^2} + V(\bm r),
\end{equation}
where $V(\bm r)$ is the electron-hole interaction potential. We recall that in two-dimensional (2D) semiconductors the interaction between charge carriers can be, as a rule, described by effective the Rytova-Keldysh potential taking into account the inhomogeneous dielectric environment~\cite{Rytova1967,1979JETPL..29..658K,Cudazzo:2011a,PhysRevB.88.045318,Durnev_2018}
\begin{equation}\label{RK:potential}
V_{RK}(r)=-\frac{\pi e^{2}}{2 r_0 }  \left[ \mathbf H_0\left( \frac{\varepsilon r}{r_0} \right) -  Y_0 \left( \frac{\varepsilon r}{r_0} \right)   \right].
\end{equation}
Here  $e$ is the electron charge,   $\mathbf H_0$ and $Y_0$ are the Struve and Neumann functions, respectively, $r_0{=2\pi \alpha/\varepsilon}$ is  a screening length in a monolayer semiconductor related to its polarizability {$\alpha$}, and $\varepsilon$ is the {(background)} dielectric constant of the environment. The potential \eqref{RK:potential} tends to the Coulomb potential 
\begin{equation}
\label{Coulomb:potential}
V_C(r) = - \frac{e^2}{\varepsilon r},
\end{equation}
at large distances $r\gg r_0$, and behaves logarithmically $\propto \ln{(r_0/r)}$ at small distances.  

In the bilayer structure the interparticle interaction potential has to be found taking into  account dielectric screening by both layers \cite{Semina:2019aa,PhysRevB.99.085108}. {The electrostatic potential $\varphi(\bm r, z)$ where $\bm r$ is the in-plane and $z$  the out-of-plane coordinate obeys the Poisson equation
   \begin{equation}
   \label{poisson}
   \Delta \varphi = -\frac{4\pi}{\varepsilon} n(\bm r,z), 
   \end{equation}
where the charge density contains the contribution of the charge, $e\delta(\bm r) \delta(z)$, and the induced charge caused by the screening, 
\[
n(\bm r) = e\delta(\bm r)\delta(z) -\bm \nabla \cdot \bm P(\bm r, z),
\]
 with 
 \begin{equation}
 \label{polarization}
 \bm P = - \sum_{i=1,2} \alpha_i \delta(z - z_i) \bm \nabla_{\bm r} \varphi(\bm r, z_i).
 \end{equation}
   being the polarization of the system. Here $\alpha_i$ is the polarizability of the corresponding layer, $\delta(z-z_i)$ ensure that the polarization is induced in the layers. We introduce $V_{11}(\bm r) = - e\varphi(\bm r, z_1)$, the potential energy of interaction where two charges are in the same layer $1$, and $V_{12}(\bm r) = -e \varphi(\bm r, z_2)$, the potential energy for the charges in different layers. These potentials can be recast as
\begin{equation}
\label{V:RK:2}
V_{11}(\bm r)  = -e \sum_{\bm q} \tilde{\varphi}_1(q) {e^{\mathrm i \bm q \bm r}}, \quad 
V_{12}(\bm r)  = -e \sum_{\bm q} \tilde{\varphi}_2(q) {e^{\mathrm i \bm q \bm r}},  
\end{equation}
where $\bm q{\parallel (xy)}$ is the two-dimensional wavevector and where we set the normalization area to unity. The corresponding Fourier transforms of the potential can be found from Eqs.~\eqref{poisson} and \eqref{polarization} as follows~\cite{Semina:2019aa}}
\begin{subequations}
\label{V:RK:999}
\begin{equation} \label{phi:intra} 
\tilde{\varphi}_1(q)=\frac{2\pi e}{\varepsilon q} 
\frac{1+qr_2(1-\xi^2)}{(1+qr_2)\left(1+qr_1-\frac{q^2r_1r_2 \xi^2}{1+q 
r_2}\right)} 
\end{equation} for  the Fourier transform of the effective potential for charge carriers in the same layer and, 
\begin{equation} 
\label{phi:inter} 
\tilde{\varphi}_2(q)=-\tilde{\varphi}_1(q)\frac{qr_1\xi}{1+qr_2}+\frac{2\pi e 
\xi}{\varepsilon q(1+qr_2)}.
\end{equation} 
\end{subequations}
 Here  
$r_1=2\pi\alpha_1{/\varepsilon}$ and  $r_2=2\pi\alpha_2{/\varepsilon}$ are effective screening radii in the first and second layer,   {and}
~$\xi=e^{-q|z_1-z_2|}=e^{-qd}$ is a dimensionless parameter depending on the interlayer distance $d=|z_1-z_2|$. The expression for the interaction of two charge carriers in layer $2$, $V_{22}(\bm r)$, can be expressed in the same way with the replacement $r_1\leftrightarrow r_2$. In the limit $d\rightarrow\infty$ the potential \eqref{phi:intra} tends to the Rytova-Keldysh potential \eqref{RK:potential} with the corresponding screening radius. For $d\rightarrow 0$ we have $V_{11}(\bm r)=V_{12}(\bm r)$, and the interaction potentials are  described by the Rytova-Keldysh expression with $r_0 = r_1 + r_2$. 

Noteworthy, the anisotropic electronic and magnetic properties of CrSBr are manifested also in the dielectric screening of the Coulomb interaction. Let us analyze this effect. For a mono- or bilayer structure, where $d$ is much smaller than the screening lengths, we can disregard the difference between the layers and solve the Poisson equation for the potential $\varphi(\bm r)$ induced by a point charge at the origin  ($\bm r=0$) with account for possible anisotropy of the dielectric constants and polarizability:
\begin{multline}
\label{Poisson:aniso}
\varepsilon_{ij} \frac{\partial^2}{\partial r_i \partial r_j} \varphi(\bm r) = -4\pi \biggl[e\delta(\bm r) \\
+ \delta(z) \alpha_{i'j'} \frac{\partial^2}{\partial r_{i'} \partial r_{j'}}\varphi(\bm r)\biggr],
\end{multline}
where $i,j=x,y,z$, and $i',j'=x,y$ denote corresponding Cartesian components, and $\varepsilon_{ij}$ is the tensor of background dielectric constants of the surrounding media (e.g., of encapsulating hexagonal boron nitride, hBN) and $\alpha_{i'j'}$ is the tensor of the mono- or bilayer in-plane polarizabilities. It is assumed that CrSBr occupies the $z=0$ plane (strictly speaking, $z=0$ corresponds to the position of a monolayer with the charge, the thickness of the monolayer is assumed to be negligible). Taking into account that $x$, $y$, and $z$ are the principal axes of the structure, we obtain the 2D Fourier transform of the potential at $z=0$ in the form
\begin{equation}
\label{RK:aniso}
\tilde \varphi(\bm q) = \frac{2\pi e}{\varepsilon_z Q + R}, 
\end{equation} 
where 
\[
Q= \sqrt{\frac{\varepsilon_x}{\varepsilon_z} q_x^2 + \frac{\varepsilon_y}{\varepsilon_z} q_y^2}, \quad R = 2\pi \left(\alpha_x q_x^2 + \alpha_y q_y^2\right).
\]
The potential energy is given by the standard expression $-e\sum_{\bm q} \tilde \varphi(\bm q)\exp{(\mathrm i \bm q\bm r)}$. In the isotropic case Eq.~\eqref{RK:aniso} is consistent with Eq.~\eqref{RK:potential}. {Interestingly, one can obtain a closed form expression for the interaction potential provided that
\begin{subequations}
\label{rescaling}
\begin{equation}
\label{special}
\frac{\varepsilon_x}{\varepsilon_y} = \frac{\alpha_x}{\alpha_y}.
\end{equation}
In such a case we recover Eq.~\eqref{RK:potential} with
\begin{equation}
\label{special:1}
\varepsilon = \varepsilon_z, \quad \mbox{and} \quad r = \sqrt{\frac{\varepsilon_z}{\varepsilon_x}x^2 +  \frac{\varepsilon_z}{\varepsilon_y}y^2},
\end{equation}
\end{subequations}
as one can readily check by performing the change of coordinates in the reciprocal space as $q_\alpha \to q_\alpha \sqrt{\varepsilon_\alpha/\varepsilon_z}$.} {While Eq.~\eqref{special} does not necessarily hold for CrSBr, we present this result for the sake of generality.}

{The microscopic approaches (see, e.g., ~\cite{PhysRevB.92.245123,2053-1583-4-2-022004,PhysRevB.98.125308}) demonstrate that the Rytova-Keldysh potential is just an approximation to the realistic electron-hole and electron-electron interaction potentials: In particular, this model does not take into account the dynamical nature of the screening~\cite{Glazov:2018aa,Scharf:2019aa} and the validity of Eqs.~\eqref{V:RK:2} and \eqref{V:RK:999} is limited to the in-plane distances exceeding the lattice constant. Still, the Rytova-Keldysh potential allows for a relatively simple and quite accurate description of the electron-hole complexes in van der Waals materials. In what follows we use the interaction given by Eqs.~\eqref{V:RK:999} in numerical calculations.}

\begin{figure}
\includegraphics[width=0.75\columnwidth]{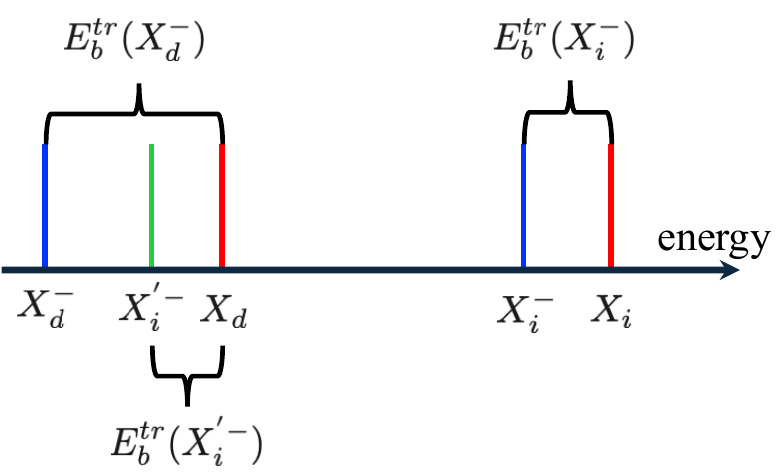}
   \caption{Schematics of spectral lines of the Coulomb complexes studied here. Energies and splittings are shown not to scale. The binding energies of trions are indicated with respect to the corresponding exciton. } 
\label{binding}
\end{figure}

Similarly, for the three-particle complex, the trion, the Hamiltonian reads
\begin{multline}
\label{H:tr}
\mathcal H_{tr} 
= 
\\
-\sum_{\alpha} \frac{\hbar^2}{2\mu_\alpha} \left[\frac{\partial^2}{\partial r_{1,\alpha}^2} + \frac{\partial^2}{\partial r_{2,\alpha}^2} + \frac{2\sigma_\alpha}{1+\sigma_\alpha} \frac{\partial}{\partial r_{1,\alpha}} \frac{\partial}{\partial r_{2,\alpha}} \right]+\\
+V_1(\bm r_1) + V_2(\bm r_2) - V_3(\bm r_1 - \bm r_2),
\end{multline}
where $\bm r_1$ and $\bm r_2$ are the distances between the first and second identical charge carrier to the non-identical one (e.g., distances between the first and second electrons and a hole for the negatively charged trion), $\sigma_\alpha$ is the ratio of the effective mass of one of the identical carriers to the effective mass of the non-identical one. The subscripts $1, 2$, and $3$ in $V(\bm r)$ distinguish the potential energies for the intra- or interlayer interaction depending on the configuration of the trion, see Fig.~\ref{fig:scheme}. For specificity, in the case of the negatively charged trions (with two electrons) the potentials $V_1$ and $V_2$ describe the attraction between each of the two electrons to the hole and the potential $V_3$ describes the repulsion between the electrons. If the charge carriers are in the same   layer one has to take $V_{11}$ \eqref{phi:intra} as potential, and for charge carriers in  different    layers one has to use $V_{12}$ \eqref{phi:inter}.

The corresponding Schr\"odinger equation $\mathcal H_{tr} \psi_{tr}(\bm r_1,\bm r_2)= \mathcal E_{tr} \psi_{tr}(\bm r_1,\bm r_2)$ allows one to determine a relative envelope function $\psi_{tr}(\bm r_1,\bm r_2)$ of the three-particle bound state and its energy $\mathcal E_{tr}$. Similarly, the exciton Schr\"odinger equation $\mathcal H_{{X}} \psi_{{X}}(\bm r)= \mathcal E_{{X}} \psi_{{X}}(\bm r)$ provides the exciton envelope and energy $\mathcal E_X$. The exciton and trion binding energies read
\begin{equation}
\label{binding:gen}
E_b^{{X}} = -\mathcal E_{{X}}, \quad E_b^{tr} = \mathcal E_{{X}} -\mathcal E_{tr}.
\end{equation}
The definitions of the binding energies are illustrated in Fig.~\ref{binding}. The situation with the direct trion $X_d^-$ is straightforward: its binding energy is reckoned from the direct neutral exciton $X_d$. For the $X_{i}^-$ trion, naturally, we have to take the energy of the indirect exciton $X_i$  as the origin of the energy. Importantly,  for $X_{i}^{'-}$ we have to take the energy of the direct exciton $X_d$ as $\mathcal E_{X}$: indeed, as soon as the energy of $X_i^{'-}$ becomes equal to the energy of the direct exciton, the trion can dissociate into a free electron and the direct exciton. It can be seen from Fig.~\ref{binding}, that, although the trion $X_{i}^{'-}$ has a lower energy than $X_{i}^{-}$, its binding energy can be smaller or even zero.  

It follows from the general analysis~\cite{Courtade:2017a} that the ground trion state (also being optically active, see Sec.~{\ref{sec:exch}} for details) corresponds to the symmetric envelope function, $\psi_{tr}(\bm r_1, \bm r_2) = \psi_{tr}(\bm r_2,\bm r_1)$. The Pauli principle thus implies that the Bloch functions of the two electrons should be different. Hence, the two electrons in a bound optically active trion should occupy either (i) different layers (in that case they can naturally be in the same type of the conduction band, $c_2$, indirect trion $X_i^{'-}$ in Fig.~\ref{fig:scheme}) or (ii) the same layer but different conduction bands, $c_1$ and $c_2$ (direct trion $X_d^{-}$ and indirect trion $X_i^{-}$ in Fig.~\ref{fig:scheme}). In the latter case, as it follows from the $\bm k\cdot \bm p$ model~\eqref{kp} the electrons have different effective masses. The analysis following Refs.~\cite{Lin:2022aa,10.1093/oxfmat/itad004} shows that in that case the Hamiltonian~\eqref{H:tr} holds but the effective masses should be properly rescaled. The estimates show that the variation of the binding energies is not very significant. The energy splitting between the $c_1$ and $c_2$ conduction bands does not affect the binding energy of the trion since it is reckoned from the energy of the bright exciton with the electron in the $c_2$ band. {As for the trions with antisymmetric envelope functions (triplet trions), $\psi_{tr}(\bm r_1, \bm r_2) =-\psi_{tr}(\bm r_2,\bm r_1)$, they are bound in two-dimensional systems with isotropic energy spectrum provided that the unpaired charge carrier ({e.g., electron in the $X^+$ trion}) is significantly lighter compared to any of the identical charge carriers ({holes in the $X^+$ trion})~\cite{Sergeev:2001aa,Courtade:2017a} or in the presence of large magnetic fields~\cite{PhysRevB.71.201312}. In one dimensional systems the triplet state is stable and all values of electron-to-hole effective mass ratio (provided that two identical charge carriers are heavier than the unpaired one or all the carriers have the same mass)~\cite{Semina:2008aa}. The detailed studies of such trions is beyond the scope of the present paper.}

We note that there are two sets of excitons and trions in bilayers, e.g., direct or intralayer exciton in the first layer and the same exciton in the second layer, the indirect (interlayer) exciton with the electron in the first layer and the indirect (interlayer) exciton with the electron in the second layer, the direct (intralayer) trion in the first layer and the same trion in the second layer, etc. For identical layers and in the absence of interlayer coupling these two sets of quasiparticles are independent but have exactly the same energies. In the presence of interlayer coupling the states are  split into layer symmetric and layer antisymmetric combinations. {The splitting is proportional to the tunneling matrix elements of electrons and holes. Owing to the antiferromagnetic order of the spins in the layers the spin states of electrons and holes in neighbouring layers are orthogonal. Hence, the tunneling is forbidden and the matrix elements vanish. Therefore, the splitting is absent if the antiferromagnetic order is not broken. That is why we neglect this splitting hereafter}.

\begin{figure*}[t]
\includegraphics[width=0.99\linewidth]{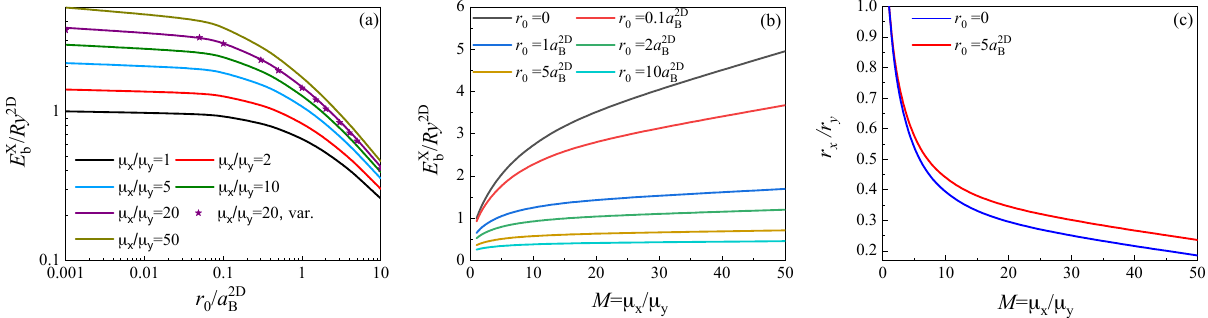}~~
   \caption{(a) Direct exciton ($X_d$) binding energy in a monolayer as function of  the screening radius numerically calculated for different anisotropies of  the reduced mass.  Stars correspond to  the variationally calculated exciton binding energy for $\mu_x/\mu_y=20$.  (b) Exciton binding energy in a monolayer as a function of  the reduced mass anisotropy, calculated for different values of the screening radius $r_0$. {(c) The ratio between  the exciton wave function dimensions along $x$- and $y$-axis, calculated for Coulomb potential and Rytova-Keldysh potential with $r_0= 5a_B$.}} 
\label{Ex_r0}
\end{figure*}


\section{Exciton and trion binding energies}\label{sec:binding}

We now turn to the results of our calculations of the exciton and trion binding energies. We consider a model case of a monolayer or bilayer encapsulated in hBN. Here we neglect the anisotropy of the polarizability of the CrSBr layers as well as of the hBN dielectric constant; estimates show that this effect is not very pronounced and does not change  the results much. {For instance, calculations in Ref.~\cite{Klein:2023aa} show that the magnitude of the dielectric susceptibility changes by about a factor of 2 depending on whether $\bm q \parallel x$ or $y$. Such anisotropy is significantly smaller than the energy dispersion anisotropy discussed above, the latter reaches $20\ldots 50$.}  We introduce  the 2D Rydberg energy and Bohr radius 
\begin{equation}\label{Bohr}
Ry^{2D}=\frac{2\mu_ye^4}{\varepsilon^2\hbar^2},~a_B^{2D}=\frac{\varepsilon\hbar^2}{2\mu_y e^2}
\end{equation}
as  the energy and distance units, with $\varepsilon$ being the background dielectric constant of hBN.

\begin{figure*}
\includegraphics[width=0.9\textwidth]{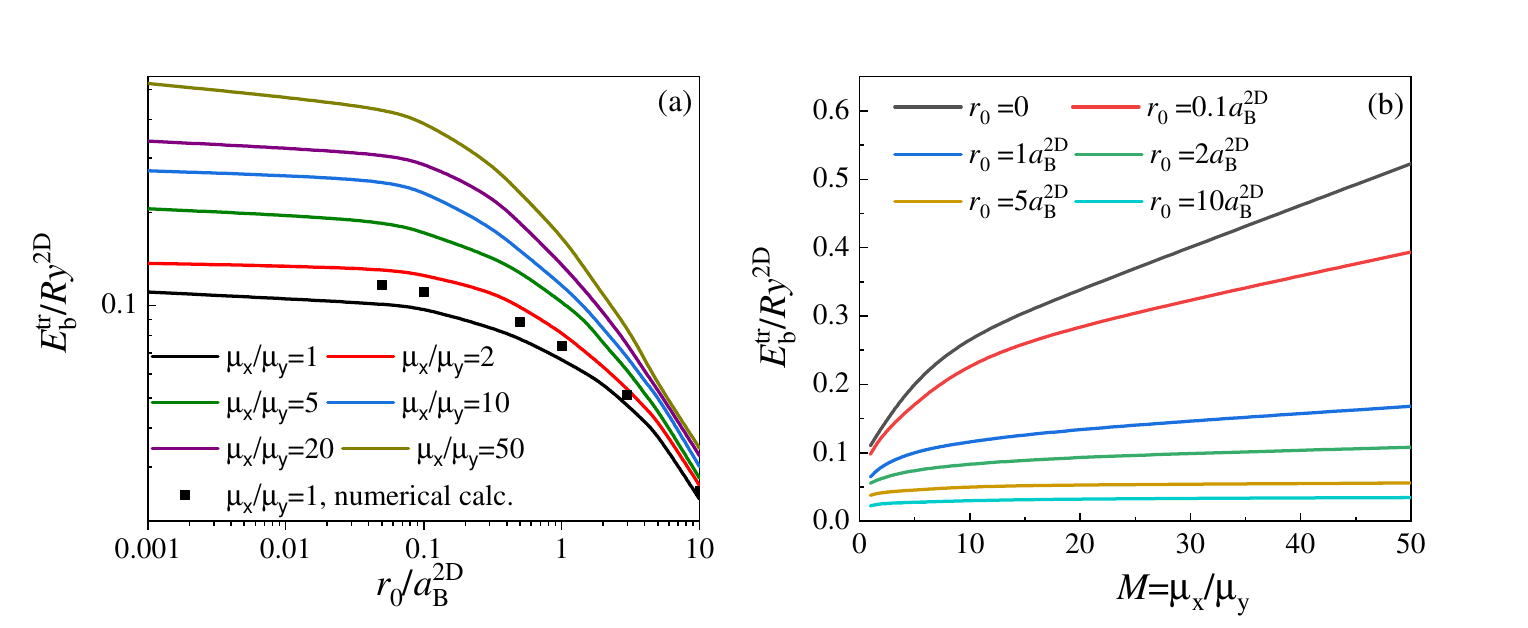}
   \caption{(a) Direct trion ($X_d^-$) binding energy in a monolayer as a function of  the screening radius calculated variationally for different anisotropies of  the reduced mass. Points show the trion binding energy for $\mu_x/\mu_y=1$ calculated using the finite element method. (b) Trion binding energy in a monolayer as a function of  the reduced mass anisotropy, calculated for different values of the screening radius $r_0$. Electron to hole mass-ratio parameters are $\sigma_x=\sigma_y=0.3$.} 
\label{Tr_r0}
\end{figure*}

\subsection{Spatially direct neutral and charged excitons}

Firstly, we consider the direct  2D  exciton and trion in a monolayer. In this case the interaction between electrons and holes is described by the Rytova-Keldysh potential \eqref{RK:potential}. To understand the effect of the anisotropy of  the effective mass  of the exciton with  the screened interaction potential, we calculated the exciton binding energy for different values of  the screening radius and reduced mass ratio along  the crystal axis $M=\mu_x/\mu_y$.  The results in 2D Bohr units, Eq.~\eqref{Bohr}, are shown in Fig. \ref{Ex_r0}.  The calculations were performed numerically by the finite element method. One can see that the effect of  the anisotropic mass on  the exciton binding energy is  strongest for  the Coulomb potential ($r_0=0$) and becomes weaker with  increasing the screening  radius. At the same time, the anisotropy of  the exciton wave function depends mainly on  the anisotropy of the reduced mass, see the Fig. \ref{Ex_r0}(c). For CrSBr with $M\approx 20$ the ratio of  the effective exciton radii along $x$ and $y$ crystal axes is about $r_x/r_y\approx 0.3$, where
\[
r_\alpha = \left(\int d^2r |\psi_x(\bm r)|^2 r_\alpha^2\right)^{1/2}.
\] For $\varepsilon=4.5$ and $r_0=3.5a_B^{2D}\approx 3.9$~nm, corresponding to the exciton binding energy in  the monolayer structure $E_B^{ex}\approx 230$~meV, we have exciton radii $r_x\approx 0.6$~nm and $r_y \approx 1.8$~nm along $x$ and $y$ axes, correspondingly. 
These estimations for the binding energies and exciton radii align with recent experiments~\cite{li2023magneticexcitonpolaritonstronglycoupled,smolenski2024largeexcitonbindingenergy}. It is noteworthy that the values are comparable with the unit lattice size, implying that the exciton can be mixed Frenkel -- Wannier-Mott type. In this regard, we stress that the results for particular CrSBr parameters should be considered as  order-of-magnitude estimates showing a possible ballpark of values.

\begin{figure}[b]
\includegraphics[width=0.99\columnwidth]{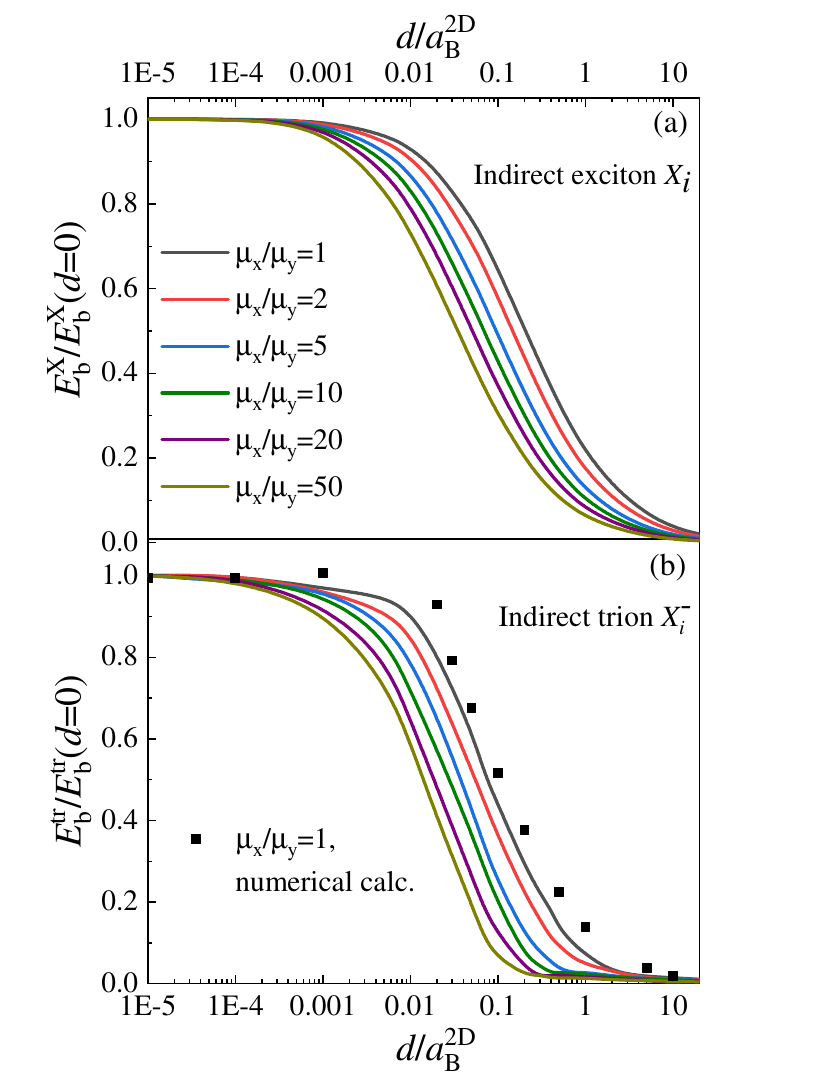}
\caption{(a) Binding energy of spatially indirect exciton $X_i$ as a function of  the interlayer distance. (b) Binding energy of spatially indirect trion $X_i^-$ (two electrons in the same layer, see Fig.~\ref{fig:scheme}). Different curves correspond to different anisotropy of  the reduced mass. Points in (b) show the trion binding energy for $\mu_x/\mu_y=1$ calculated using the finite element method. The color code for the curves is the same for all panels. Here, the Coulomb potential was used. Other parameters are the same as in Fig.~\ref{Tr_r0}.} \label{Exciton_Coulomb_masses}
\end{figure}

Direct trion binding energies were calculated by  the variational method. To provide the uniformity of the approach and improve  the numerical accuracy we also employed variational calculation of the exciton. We use the following trial function for the exciton
\begin{equation}\label{Ex_trial}
\Psi_{X}(x,y,\alpha,\beta)=\exp\left(-\alpha  \sqrt{x^2+\beta ^2 y^2}\right),
\end{equation}
with two trial parameters $\alpha$ and $\beta$.  The parameter $\alpha$ describes the attraction between  the electron and  the hole, and, consequently, the exciton radius.  The parameter $\beta$ describes the anisotropy of  the exciton wave function. Comparison with the finite element method shows that the accuracy of such a wavefunction is better than 4$\%$,  see Fig. \ref{Ex_r0} (a),  where the stars show the results of variational calculation for $\mu_x/\mu_y=20$.  Correspondingly, for the trion we use the  trial function based on the exciton trial function~\eqref{Ex_trial} and extending the well-established variational functions to the anisotropic mass case~\cite{Sergeev:2001aa,Courtade:2017a,Semina_2022,10.1093/oxfmat/itad004}
\begin{multline}\label{Tr_trial}
 \Psi_{tr}(x_1,y_1,x_2,y_2,\alpha_1,\alpha_2,\beta_1,\beta_2{, \gamma}) \\=
   \left[\Psi_{{X}}(x_1,y_1,\alpha_1,\beta_1)\Psi_{{X}}(x_2,y_2,\alpha_2,\beta_2)\right. \\ +
 \left.\Psi_{{X}}(x_2,y_2,\alpha_1,\beta_1)\Psi_{{X}}(x_1,y_1,\alpha_2,\beta_2)  \right]\\
 \\\times \left(1+\gamma  \sqrt{(x_1-x_2)^2+(y_1-y_2)^2}\right).
\end{multline}
Here $\alpha_1$, $\alpha_2$, $\beta_1$, $\beta_2$ and $\gamma$ are trial parameters.  The parameters  $\alpha_1$, $\alpha_2$, $\beta_1$, $\beta_2$ have a similar meaning as  the parameters $\alpha$ and $\beta$  for the exciton,  the parameter $\gamma$ allows one to optimize the trion wave function to reduce  the repulsion between two identical charge carriers. Note that one can further improve  the trial function \eqref{Ex_trial} by introducing an additional parameter $\delta$,  and, thus, allowing anisotropy in the correlation term $\left(1+\gamma  \sqrt{(x_1-x_2)^2+\delta^2(y_1-y_2)^2}\right)$. Our estimates show that it does not improve  the accuracy of  the calculations much even in the case of the strong anisotropic, cf. the one-dimensional case, Ref.~\cite{Semina_2022} and Appendix~\ref{AA}, but makes them computationally more demanding.  The accuracy of such a variational approach has been recently estimated in Ref.~\cite{kumar2024trionsmonolayertransitionmetal} by comparison with more sophisticated calculations. To provide an additional check of the accuracy we have performed the finite element method calculations of the trion binding energies for the isotropic masses, see black points in Fig.~\ref{Tr_r0}(a) for the spatially direct trion and in Fig.~\ref{Exciton_Coulomb_masses}(b) for the indirect trion. Our estimations show that the accuracy of the variational calculation of the absolute trion energy is better than 5-10\% range. For the binding energy the accuracy is somewhat lower, 20-30 \%, because it is obtained as a difference of two large (on the order of exciton binding energy) values.

Figure~\ref{Tr_r0} shows the dependence of  the trion binding energy on the mass anisotropy and screening radius. The shown dependences are very similar to  the corresponding dependences for  the exciton, although with smaller values of the binding energy. Roughly, the direct trion binding energy is about $10\%$ of the exciton binding energy being consistent with other two-dimensional systems~\cite{Sergeev:2001aa,PhysRevB.72.035332,Courtade:2017a,PhysRevB.96.035131,kumar2024trionsmonolayertransitionmetal}. It is noteworthy that our results on the direct exciton and trion binding energies and anisotropy as functions of the mass ratio are in line with previous works on other anisotropic two-dimensional materials~\cite{PhysRevB.93.115314,PhysRevB.98.235401}.

\subsection{Spatially indirect excitons and trions}

\begin{figure*}
\includegraphics[width=0.99\textwidth]{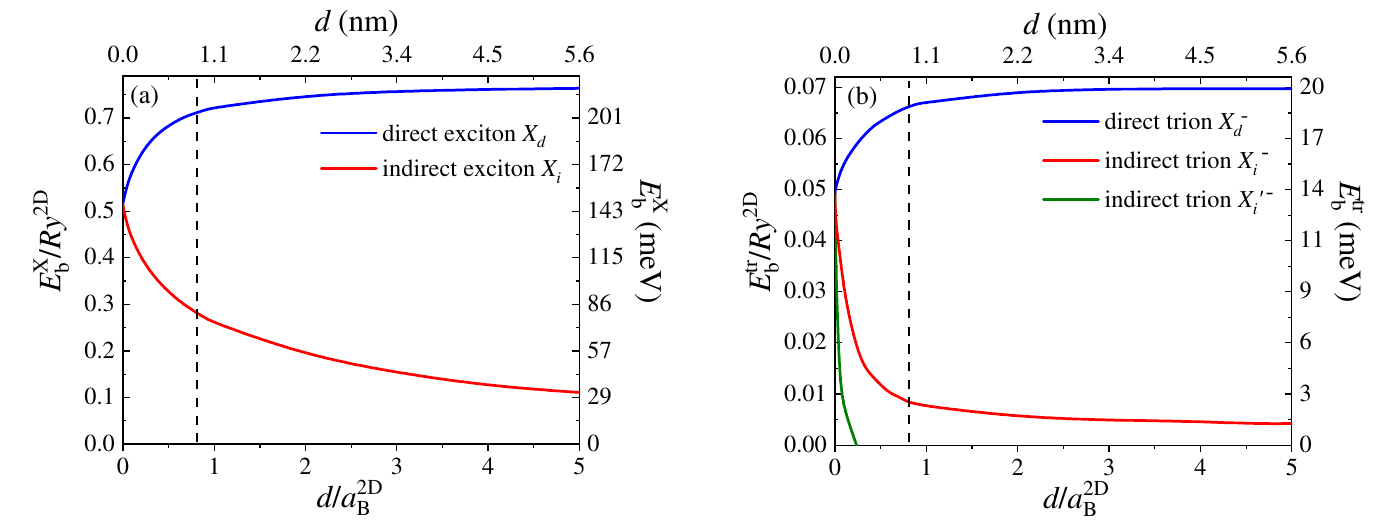}
   \caption{Binding energies of direct and spatially indirect excitons (a) and trions  (b) calculated using the screened Rytova-Keldysh potential  and background dielectric constant $\varepsilon=4.5$, $r_0=3.9$~nm. Effective masses are $m_x^e=7.31 m_0$, $m_y^e=0.14 m_0$, $m_x^h=2.84 m_0$, $m_y^h=0.45 m_0$.  Dashed vertical lines in panels  (b) and (d) correspond to interlayer distance  $d=0.8$~nm.  }\label{Eb_ind} 
\end{figure*}

Let us start with the analysis of the  effective mass anisotropy effect on spatially indirect complexes. For the purpose of illustration, in calculations shown in Fig.~\ref{Exciton_Coulomb_masses} we took the Coulomb potential to describe  the interaction between   the charge carriers.  For   the screened  potential \eqref{V:RK:2} the results are qualitatively the same, although quantitatively different, see below.

Panels (a) and (b) of Fig.~\ref{Exciton_Coulomb_masses}  show, respectively, the interlayer distance $d$ dependence of  the binding energy of the spatially indirect exciton $X_i$ and the spatially indirect trion $X_i'$ with two electrons in the same layer, see Fig.~\ref{fig:scheme} for details. The binding energies are normalized to their values for the direct complexes ($d=0$). The calculations were performed by the variational method with the trial functions \eqref{Ex_trial} and \eqref{Tr_trial}. The binding energies decrease with  the increase in the interlayer distance because the attraction of the charge carriers becomes weaker~\cite{Sergeev2003,Semina:2019aa}.
One can clearly see, that the larger the value of the mass anisotropy $M=\mu_x/\mu_y$ is, the steeper is the decrease of the exciton binding energy with increasing distance $d$. Similar behavior is observed for the trion $X_i^-$. Note, that too small values of $d$, $d\lesssim 0.1 a_B^{2D}$, are typically unrealistic and we show them in the figure to illustrate the correct asymptotics. Noteworthy, for very large $d>d_{cr} \sim 30 a_B^{2D}$ the indirect trion binding energy becomes negative meaning that the trion becomes unbound~\cite{Sergeev2003}, while the interlayer exciton is bound for any $d$~\cite{2011JETPL..94..574S}. We recall that the trion binding energy is defined as a difference of the energy positions of the exciton and trion lines in the optical spectra in the limit of vanishing electron density, see Fig.~\ref{binding}. For the negative binding energy the state of the exciton and unbound electron has a lower energy compared to the energy of the trion, it means that the trion becomes unstable.  Our calculation shows, that for $X_i^{'-}$  the binding energy decreases even more rapidly and becomes zero at unrealistically small values of $d\sim 0.1a_B^{2D}$, see below for details.

The main results of this Section are summarized in Fig.~\ref{Eb_ind} where the calculated binding energies of the spatially direct and indirect excitons $X_d$ and $X_i$ as well as of all three trions $X_d^-$, $X_i^-$, and $X_i^{'-}$ are shown. These calculations were performed with the screened potential \eqref{V:RK:2}. In these calculations we set the mass-ratio parameters $\sigma_x=2.7$  and $\sigma_y=0.3$ in Eq.~\eqref{H:tr} in accord with Ref.~\cite{Klein:2023aa}, see details in the caption. The results are presented both in the Bohr [panels (a) and (c)] and dimensional [panels (b) and (d)] units, roughly corresponding to the hBN encapsulated CrSBr bilayer structure with the background dielectric constant  $\varepsilon=4.5$, screening radii $r_1=r_2=r_0=3.5 a_B^{2D}\approx 3.9$~nm, and the effective masses  $m_x^e=7.31 m_0$, $m_y^e=0.14 m_0$, $m_x^h=2.84 m_0$, $m_y^h=0.45 m_0$. For this set of parameters we have $Ry^{2D}=286$~meV and $a_B^{2D}=1.12$~nm and the binding energy of the direct exciton at $d=0$ is about $150$~meV.

We stress that here both the interlayer distance $d$ and  the effective screening radius $r_0$ should be treated as effective, phenomenological quantities. The model used  is simpler than reality in several aspects, particularly, because  the interlayer distance is on the order of  the unit cell size in realistic bilayer structures. It implies that the effective mass approximation and macroscopic description of the screening by a dielectric between the layers are not entirely applicable.

One can see from Fig.~\ref{Eb_ind} (a) that  for the indirect exciton $X_i$,  the binding energy  decreases with increasing  interlayer distance, while  for the direct exciton  $X_d$  the binding energy increases. The latter is  the result of diminishing screening by the second layer in the bilayer. For $d\to 0$ the spatially direct exciton binding energy corresponds to the value in a monolayer with  twice  as large screening radius, $r_0=7 a_B$. At $d\to \infty$ the  screening by the second layer becomes suppressed and the direct exciton binding energy increases up to the value in an isolated monolayer with $r_0=3.5 a_B$~\cite{Semina:2019aa}). The decrease of the interlayer, spatially indirect, exciton binding energy is mainly caused by the decrease of the Coulomb interaction, while the suppression of  the screening plays a minor role. 

The same tendencies hold also for the trions, see Fig.~\ref{Eb_ind} (b).  For the spatially direct trion $X_d^-$,  the binding energy is about $14\ldots 20$ meV,  in reasonable agreement with the measurements reported in Ref. \cite{Tabataba-Vakili:2024aa}.  The binding energy of indirect trions decreases with increasing  interlayer distance. The effect is most pronounced for the $X_i^{'-}$ whose binding energy drops to negligible values at $d\gtrsim 0.25 a_B^{2D}$. Such a trion  is unlikely to be observed in bilayer CrSBr. The binding energy of the indirect trion $X_i^-$ also decreases with increasing $d$, but less steeply. For realistic $d=0.7\ldots 0.9$~nm its binding energy is about $2\ldots 3$~meV. Such a state can, in principle, be observed in experiments on bilayer CrSBr. Since its binding energy is significantly smaller than that reported in Ref. \cite{Tabataba-Vakili:2024aa}, we can conclude, that the most likely type of the trion observed in Ref.~\cite{Tabataba-Vakili:2024aa} is the direct intralayer one with a binding energy in the range $18 \ldots  20$~meV for the relevant interlayer distance $d$.

\section{Exciton radiative decay and electron-hole exchange interaction}\label{sec:exch}

The electron-hole exchange interaction plays a key role in determining the fine structure of  Coulomb complexes~\cite{ivchenko05a}. Particularly, the long-range exchange interaction related to virtual annihilation of an exciton and its subsequent creation controls the splitting of the exciton radiative doublet in conventional semiconductor quantum well structures~\cite{maialle93,goupalov98} and in transition-metal dichalcogenide monolayers~\cite{Yu:2014fk-1,glazov2014exciton,PhysRevB.89.205303}.

Let us analyze the effects of the long-range exchange interaction on excitonic complexes in two-dimensional CrSBr. First, following Refs.~\cite{ivchenko05a,glazov2014exciton,PhysRevLett.123.067401} we calculate the bright intralayer exciton radiative decay rate as
\begin{equation}
\label{gamma0}
\Gamma_0 \equiv \Gamma_0^{yy} = \frac{2\pi q_b}{\varepsilon_b \hbar} \left(\frac{e |p_{cv}\psi_{X}(0)|}{m_0\omega_0} \right)^2,
\end{equation}
where $q_b = \omega_0 \sqrt{\varepsilon_b}/c$ is the wavevector of a photon with the exciton resonance frequency $\omega_0 = (E_g - E_b^{X})/\hbar$ and  $\varepsilon_b$ is the background dielectric constant of the surrounding medium taken at the frequency $\omega_0$, the interband momentum matrix element $p_{cv}$ was introduced in Eq.~\eqref{pcv} and $\psi_{X}(0)$ is the electron-hole relative motion wavefunction taken at the coinciding coordinates of the particles. We stress that the optical transition occurs only in the $y$-polarization. The optical transitions related to an interlayer exciton are forbidden in this model because  interlayer tunneling is absent. It can be induced by an external magnetic field that results in  spin canting~\cite{Wilson:2021aa,Tabataba-Vakili:2024aa}. In that case, the interlayer exciton acquires an oscillator strength due to the field-induced mixing with the intralayer one.

As a next step we evaluate the exciton self-energy due to the virtual emission and absorption of photons. It follows from Refs.~\cite{glazov2014exciton,Iakovlev:2024aa} that
\begin{equation}
\label{shift}
\Delta E (\bm K)= \hbar\Gamma_0 \frac{K_y^2}{q_b\sqrt{K_x^2+K_y^2} },
\end{equation}
where $\bm K = (K_x,K_y)$ is the exciton in-plane wavevector and Eq.~\eqref{shift} holds for $K\gg q_b$, while for $K\leqslant q_b$ the exciton  decays radiatively. Unlike the fine structure splitting of excitons in the longitudinal and transversal states in transition metal dichalcogenides, the energy of the only optically active state in CrSBr acquires an anisotropic correction [cf. Ref.~\cite{PhysRevLett.132.126902}]:
\begin{equation}
\label{X:disper}
E(\bm K) = \frac{\hbar^2 K_x^2}{2(m^e_x+m^h_x)} + \frac{\hbar^2 K_y^2}{2(m^e_y+m^h_y)} + \Delta E(\bm K).
\end{equation}
The anisotropy of the dispersion for excitons stems both from the anisotropic dispersion of free electrons and holes and from the long-range exchange interaction~\eqref{shift}. The presence of such contribution, $\Delta E(\bm K)$, results in the anisotropic enhancement of the exciton group velocity $\bm v_g = \hbar^{-1} \nabla_{\bm K} E(\bm K)$~\cite{glazov2024ultrafastexcitontransportvan} with $v_{g,y}(\bm K) > v_{g,x}(\bm K)$, and can, in principle, be explored in exciton propagation experiments. Similarly, an additional anisotropic contribution to the Fermi polaron/Suris tetron dispersion appears as a result of the long-range exchange interaction~\cite{Iakovlev:2024aa}. It is described by Eq.~\eqref{shift} with the replacement of the exciton radiative decay rate $\Gamma_0$ by the Fermi polaron decay rate $\Gamma_{FP} =f \Gamma_0 \propto \Gamma_0 (E_F/E_{b}^{tr})$, see Sec.~\ref{sec:FP} and Eq.~\eqref{f:suris:tetron} for details.

\section{Conclusion}\label{sec:concl}

To conclude, we have presented the calculations of excitons and  trions in two-dimensional semiconductors with anisotropic energy dispersion. Combining the finite element method and variational calculations we have studied the trends of the spatially direct and indirect exciton and trion binding energies as functions of the effective mass anisotropy, screening, and interlayer distance in bilayer structures. Estimates for bilayer CrSBr yield the binding energy of intralayer excitons and trions to be of about $200$~meV and $18\ldots 20$~meV, respectively. These values are in line with recent experiments. In two-layer structures the binding energies of spatially direct excitons and trions increase, while the binding energies of the interlayer, spatially indirect excitons and trions decrease with  increasing  interlayer distance. 

We have also briefly analyzed the effects of correlations between photoexcited exciton and resident electrons which give rise to Fermi polarons also known as Suris tetrons quasiparticles, and addressed the effects of  light-matter coupling, including radiative decay and fine structure splittings.

The anisotropic energy spectrum of free charge carriers and anisotropic selection rules for optical transitions result in anisotropic dispersion of Coulomb complexes at finite momenta. It is expected that the excitons, trions, and Fermi polarons/Suris tetrons have higher propagation velocities along the magnetic easy $b$-axis compared to the velocities for propagating along the intermediate  $a$-axis.

\acknowledgements

M.A.S. and M.M.G. work (theory) has been supported by the Russian Science Foundation grant 23-12-00142. F.~T.-V., A.~B. and A.~H. acknowledge funding by the Deutsche Forschungsgemeinschaft (DFG, German Research Foundation) within Germany's Excellence Strategy under grant No.~EXC-2111-390814868 (Munich Center for Quantum Science and Technology, MCQST). F.~T.-V. also acknowledges funding by the European Union's Framework Programme for Research and Innovation Horizon Europe under the Marie Sklodowska-Curie Actions grant agreement No.~101058981. A.~R. acknowledges funding by the Bavarian Hightech Agenda within the Munich Quantum Valley doctoral fellowship program.

\appendix
\setcounter{figure}{0}
\renewcommand{\thefigure}{A\arabic{figure}}

\section{Applicability of quasi-1D approximation}
 \label{AA}

 \begin{figure*}
\includegraphics[width=0.99\textwidth]{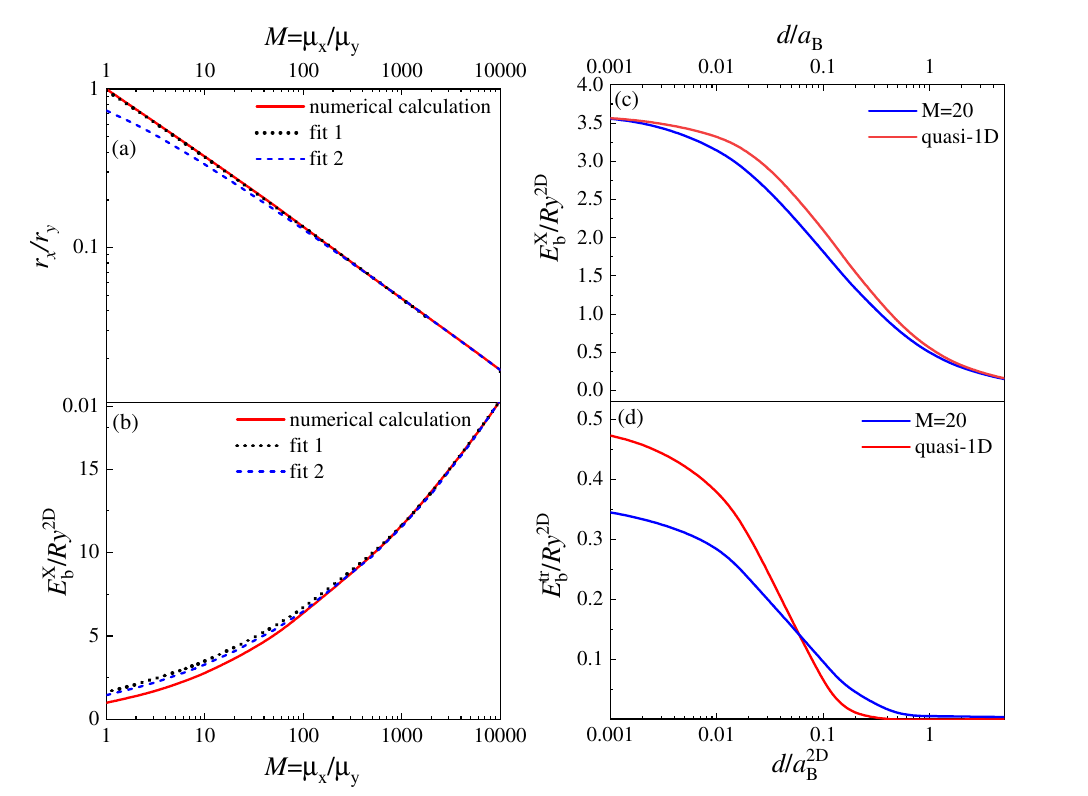}
   \caption{(a) Ratio between the exciton radii along the crystal axis $R=r_x/r_y$. (b) Binding energy of the exciton ground state with Coulomb interaction and anisotropic reduced mass. Fit 1 and fit 2 correspond to Eqs.~\eqref{fit11} and \eqref{fit22}, respectively. Inset shows the zoom in the region of $M\approx 1$. Panels (c) and (d) show the binding energies of the spatially indirect  exciton (c) and trion (d) with Coulomb interaction calculated in quasi-1D approximation and for 2D with $M=\mu_x/\mu_y=20$.}
\label{Quasi_1D}
\end{figure*}

In the case of different reduced masses $\mu_x\neq \mu_y$ the 2D exciton ground state wave function would not be axially-symmetric.  To illustrate this, we calculate the ratio between the exciton radii along the crystal axis $R=r_x/r_y$ for  the exciton with Coulomb interaction between charge carriers as a function of the ratio between exciton reduced masses $M=\mu_x/\mu_y$. The results are shown in Fig.~\ref{Quasi_1D} (a) by the solid line. For CrSBr, $M\approx 20$ and we obtain $R\approx 0.28$, which is slightly smaller than for exciton with the Rytova-Keldysh potential. In the  limit $M\rightarrow \infty$ we derived  two asymptotics. The first one is:
\begin{equation}\label{fit1}
R(M)=B_1\sqrt{\frac{F(A_1M)}{A_1M}},
\end{equation}
where $A_1$ and $B_1$ are fitting parameters, $F(x)$ is the solution of the equation for $w$ $x=we^w$. The corresponding fit is shown in Fig.~\ref{Quasi_1D}(a) by the dotted line. The second fit is  simpler:
\begin{equation}\label{fit2}
R(M)=B_2\sqrt{\frac{\ln{(A_2M)}}{A_2M}},
\end{equation}
with $A_2$ and $B_2$ being fitting parameters. The result of this fit is shown by the dashed line in  Fig.~\ref{Quasi_1D}(a). The first fit is almost ideal, the second one is good if $M\gtrsim 100$.

In Fig. \ref{Quasi_1D} (b) the  dependence of the exciton ground state  binding energy with Coulomb interaction and anisotropic reduced mass, $E_b^{ex}$, calculated numerically as a function of $M=\mu_x/\mu_y$ is shown  by  the solid line. As for the parameter $R$ we also derived two asymptotes. The first is:
\begin{equation}\label{fit11}
E_b^{ex}=B_3 F^4(A_3M),
\end{equation}
with $A_3$ and $B_3$ being fitting parameters. The results of the fitting with Eq. \eqref{fit11} is shown in Fig. \ref{Quasi_1D}(b) by the dotted line. The second asymptotic is
\begin{equation}\label{fit22}
E_b^{ex}=B_4 \ln^4(A_4M), 
\end{equation}
with $A_4$ and $B_4$ being fitting parameters. The result of the fitting with Eq. \eqref{fit22} is shown in Fig. \ref{Quasi_1D}(b) by the dashed line. Both fits demonstrate very similar results and are applicable if $M\gtrsim 100$. The results, shown in Fig. \ref{Quasi_1D}(b) also demonstrate that for CrSBr with $M\approx 20$ the strict 1D approximation is not applicable.  For the screened Coulomb interaction the asymtotics would be the same but with different constants.

In a quasi-1D system the charge carriers can move in an infinite stripe with small thickness $a$. In that case, the Coulomb potential can be approximated by the simple expression:
\begin{equation}
V(r)=-\frac{e^2}{\varepsilon \sqrt{r^2+a^2}},
\end{equation} 
where $\varepsilon$ is the dielectric constant of the environment, $r$ is the distance between the electron and hole. For small  values of $a$
the binding energy of an exciton $E_{ex}^B$ with Coulomb interaction  diverges logarithmically as $E_{ex}^{B}\propto \ln^2\left(\frac{a_B}{a}\right)$ \cite{loudon59}.   
If we consider the bilayer system with identical layers, the model described above is valid only if the interlayer distance $d$ is larger than $a$, $d>a$.

Here we compare the results for indirect excitons and trions with Coulomb interaction calculated in 2D with anisotropic mass  and in quasi-1D approximation. We took the ratio of the reduced masses along $x$ and $y$ coordinates $M=\mu_x/\mu_y=20$ similar to the one reported for CrSBr and the value of $a$ to match the binding energy of 2D and quasi-1D direct excitons. In  Fig. \ref{Quasi_1D}(c) we show the binding energies of quasi-1D and 2D excitons with Coulomb interaction as a function of the interlayer distance in Bohr units. One can see  very good agreement. In  Fig. \ref{Quasi_1D}(d) a similar comparison for the trion of the indirect exciton $X^-_i$ is shown. Unlike excitons, for trions there is no agreement at all. The binding energy of the  trion  within the quasi-1D approximation is much larger in the limit $d\rightarrow \infty$, and it decreases with the interlayer distance much faster. So, for trions the quasi-1D approximation is not well applicable.

\section{Correlations with the Fermi sea}\label{sec:FP}

While the trion approach is convenient in many cases, particularly for analyzing the binding energies and wavefunctions, in structures with free charge carriers an exciton can bind with any of  the resident electrons leaving a hole in a Fermi sea. The many-body correlations call for  a special analysis, usually carried out within the Fermi polaron/Suris tetron approach~\cite{suris:correlation,PhysRevA.85.021602,Sidler:2016aa,PhysRevB.95.035417,PhysRevX.9.041019,Glazov:2020wf,Iakovlev_2023}. Here we focus on the simplest possible but analytical description valid for not too high electron densities where the Fermi energy is small compared to the trion binding energy. In this case we obtain simple expressions that can be used for fitting the experimental data.

For the sake of simplicity we consider in this Section the exciton as a rigid quasiparticle and introduce a short-range interaction constant $V_0<0$ describing the exciton-electron attraction (the specifics of long-range interactions in the Fermi polaron problem was studied in Ref.~\cite{PhysRevB.109.214208}). The electron-exciton scattering amplitude $T(E, \bm q)$ can be expressed as~\cite{suris:correlation,PhysRevA.85.021602,PhysRevB.95.035417,PhysRevX.9.041019,Glazov:2020wf}
\begin{equation}
\label{T:ampl}
T(E,\bm q) =  \frac{V_0}{1-V_0 S(E,\bm q)}, 
\end{equation}
with
\begin{equation}
\label{S:sum}
 S(E,\bm q) = \sum_{\bm p} \frac{1-n_{\bm p}}{E - E_{X}({\bm q - \bm p}) - E_e(\bm p) + E_e(\bm q)},
\end{equation}
where $E_e(\bm k)$ and $E_{X}(\bm k)$ are, respectively, the electron and the exciton dispersions taken in the effective mass approximation, cf. Eqs.~\eqref{masses}, and $n_{\bm p}$ is the electron distribution function. Hereafter, we consider the zero  temperature  limit, hence, $n_{\bm p}$ is a step function. The exciton self-energy with allowance for the exciton-electron scattering can be recast as (we consider  the exciton at rest)~\cite{suris:correlation,PhysRevA.85.021602,PhysRevB.95.035417,PhysRevX.9.041019,Glazov:2020wf} (see also \cite{PhysRevB.98.235203,PhysRevB.102.085304} for similar but alternative approaches)
\begin{equation}
\label{sigma:exciton}
\Sigma_{X}(E) = \sum_{\bm q} n_{\bm q} T(E,\bm q).
\end{equation}
Accordingly, the poles of the exciton Green's function
\begin{equation}
\label{exciton:Greens}
\mathcal G_{ex}(E) = \frac{1}{E - \Sigma_{X}(E) + \mathrm i \gamma_{X}},
\end{equation}
determine the correlated Fermi polaron states. Here $\gamma_{X}$ is the exciton decay rate.

At $\bm q=0$ the poles of $T$ correspond to the bound exciton-electron state, i.e., to the trion. Unlike two-dimensional systems with  isotropic dispersion, here the difference of the effective masses along the principal axes results in the specifics of the scattering problem. To analyze it let us neglect the occupancy of electronic states by setting $1-n_{\bm p} = 1$ in the summation in Eq.~\eqref{S:sum}, and recast the equation for the pole of $T$ in the form
\begin{equation}
\label{T:pole}
1 = \frac{V_0}{(2\pi)^2} \int_{-Q_x}^{Q_x} dk_x \int_{-Q_y}^{Q_y} dk_y \frac{1}{E - \frac{\hbar^2 k_x^2}{2\mu_x^{eX}} -  \frac{\hbar^2 k_x^2}{2\mu_y^{eX}}},
\end{equation}
where $\mu_\alpha^{eX}$ ($\alpha=x$ or $y$) are the electron-exciton reduced masses and  the cut-off wavevectors $Q_\alpha$ are introduced to account for a finite spread of the exciton-trion interaction potential~\cite{Glazov:2020wf}. Taking into account that {$\mu_y^{eX} \ll \mu_x^{eX}$} one can integrate over $k_y$ first by extending the integration limits $Q_y \to \infty$. The remaining integral is trivially evaluated with the result
\begin{equation}
\label{T:pole:1}
-\frac{1}{V_0 \mathcal D } =  \ln{\left(1+ 2\frac{E_Q+\sqrt{-EE_Q + E_Q^2}}{-E} \right)},
\end{equation}
where $\mathcal D = \sqrt{\mu_x^{eX}\mu_y^{eX}}/(2\pi \hbar^2)$ and $E_Q = \hbar^2 Q_x^2/2{\mu_x^{eX}}$. Hence, the binding energy is related to the interaction parameter $V_0$ as
\begin{multline}
\label{T:pole:2}
E_b^{tr} = - E= 4E_Q \frac{e^{1/V_0\mathcal D}}{(1-e^{1/V_0\mathcal D})^2}\\
 =4E_Q
\begin{cases}
e^{1/V_0\mathcal D}, \quad |V_0\mathcal D| \ll 1,\\
(V_0 \mathcal D)^2, \quad |V_0\mathcal D| \gg 1.
\end{cases}
\end{multline}
The former situation, $|V_0 \mathcal D| \ll 1$, corresponds to a widely-studied two-dimensional case where the binding energy is exponentially small as compared to $E_Q$. In such a case, $4E_Q$ plays the role of the cut-off energy. The latter situation, $|V_0\mathcal D| \gg 1$, corresponds to the quasi-1D limit where the binding energy of a state is a short-range potential that scales as the squared depth of the potential~\cite{ll3_eng} (see also Refs.~\cite{McGuire:1966aa,PhysRevA.84.033607} where the Fermi polarons in one dimension were studied in detail).

The simplified analysis of the Fermi polaron/Suris tetron states can be done following Refs.~\cite{Glazov:2020wf,Iakovlev_2023}. Here we outline the simplest ``trion-pole'' approximation where for $E \approx -E_b^{tr}$ the scattering amplitude is approximated as
\begin{equation}
\label{trion:pole}
T(E,\bm q) \approx \frac{\mathcal D^{-1} E_b^{tr}\sqrt{1+ E_b^{tr}/E_{Q}}}{E + \frac{\hbar^2 q_x^2}{2m_x} + \frac{\hbar^2 q_y^2}{2m_y}+E_b^{tr}},
\end{equation}
where $1/m_\alpha = 1/m_\alpha^e - 1/m_\alpha^{tr}$ and $m_\alpha^{tr}$ is the trion effective mass. Evaluating the self-energy~\eqref{sigma:exciton} and solving for the poles of the exciton Green's function~\eqref{exciton:Greens} we obtain the Fermi polaron energies~\footnote{For significant $V_0\mathcal D$ a simple expression~\eqref{sigma:exciton} for the exciton self-energy may not be applicable due to strong interaction of the Fermi sea hole with the exciton.}. In particular, the attractive Fermi polaron, also known as Suris tetron, energy reads
\begin{equation}
\label{suris:tetron}
E = -E_{b}^{tr} - \mathcal B E_F \sqrt{1+ E_b^{tr}/E_{Q}},
\end{equation}
where $E_F$ is the electron Fermi energy and $\mathcal B$ is a coefficient of the order of unity that depends on the effective masses. Eq.~\eqref{suris:tetron} holds for $E_F \ll E_b^{tr}$. Accordingly, the exciton weight in the Fermi-polaron wavefunction is given by
\begin{equation}
\label{f:suris:tetron}
f = \mathcal A \frac{E_F}{E_b^{tr}} \sqrt{1+ E_b^{tr}/E_{Q}},
\end{equation}
with $\mathcal A \sim 1$ being another mass dependent coefficient. The quantity $f$ describes, in particular, the redistribution of the oscillator strength between excitons and polarons.
This analysis can be extended to account for the direct and indirect excitons and the corresponding trions, justifying the approach used in Ref.~\cite{Tabataba-Vakili:2024aa}.


\begin{thebibliography}{58}%
\makeatletter
\providecommand \@ifxundefined [1]{%
 \@ifx{#1\undefined}
}%
\providecommand \@ifnum [1]{%
 \ifnum #1\expandafter \@firstoftwo
 \else \expandafter \@secondoftwo
 \fi
}%
\providecommand \@ifx [1]{%
 \ifx #1\expandafter \@firstoftwo
 \else \expandafter \@secondoftwo
 \fi
}%
\providecommand \natexlab [1]{#1}%
\providecommand \enquote  [1]{``#1''}%
\providecommand \bibnamefont  [1]{#1}%
\providecommand \bibfnamefont [1]{#1}%
\providecommand \citenamefont [1]{#1}%
\providecommand \href@noop [0]{\@secondoftwo}%
\providecommand \href [0]{\begingroup \@sanitize@url \@href}%
\providecommand \@href[1]{\@@startlink{#1}\@@href}%
\providecommand \@@href[1]{\endgroup#1\@@endlink}%
\providecommand \@sanitize@url [0]{\catcode `\\12\catcode `\$12\catcode
  `\&12\catcode `\#12\catcode `\^12\catcode `\_12\catcode `\%12\relax}%
\providecommand \@@startlink[1]{}%
\providecommand \@@endlink[0]{}%
\providecommand \url  [0]{\begingroup\@sanitize@url \@url }%
\providecommand \@url [1]{\endgroup\@href {#1}{\urlprefix }}%
\providecommand \urlprefix  [0]{URL }%
\providecommand \Eprint [0]{\href }%
\providecommand \doibase [0]{https://doi.org/}%
\providecommand \selectlanguage [0]{\@gobble}%
\providecommand \bibinfo  [0]{\@secondoftwo}%
\providecommand \bibfield  [0]{\@secondoftwo}%
\providecommand \translation [1]{[#1]}%
\providecommand \BibitemOpen [0]{}%
\providecommand \bibitemStop [0]{}%
\providecommand \bibitemNoStop [0]{.\EOS\space}%
\providecommand \EOS [0]{\spacefactor3000\relax}%
\providecommand \BibitemShut  [1]{\csname bibitem#1\endcsname}%
\let\auto@bib@innerbib\@empty
\bibitem [{\citenamefont {Telford}\ \emph {et~al.}(2020)\citenamefont
  {Telford}, \citenamefont {Dismukes}, \citenamefont {Lee}, \citenamefont
  {Cheng}, \citenamefont {Wieteska}, \citenamefont {Bartholomew}, \citenamefont
  {Chen}, \citenamefont {Xu}, \citenamefont {Pasupathy}, \citenamefont {Zhu},
  \citenamefont {Dean},\ and\ \citenamefont {Roy}}]{Telford:2020aa}%
  \BibitemOpen
  \bibfield  {author} {\bibinfo {author} {\bibfnamefont {E.~J.}\ \bibnamefont
  {Telford}}, \bibinfo {author} {\bibfnamefont {A.~H.}\ \bibnamefont
  {Dismukes}}, \bibinfo {author} {\bibfnamefont {K.}~\bibnamefont {Lee}},
  \bibinfo {author} {\bibfnamefont {M.}~\bibnamefont {Cheng}}, \bibinfo
  {author} {\bibfnamefont {A.}~\bibnamefont {Wieteska}}, \bibinfo {author}
  {\bibfnamefont {A.~K.}\ \bibnamefont {Bartholomew}}, \bibinfo {author}
  {\bibfnamefont {Y.-S.}\ \bibnamefont {Chen}}, \bibinfo {author}
  {\bibfnamefont {X.}~\bibnamefont {Xu}}, \bibinfo {author} {\bibfnamefont
  {A.~N.}\ \bibnamefont {Pasupathy}}, \bibinfo {author} {\bibfnamefont
  {X.}~\bibnamefont {Zhu}}, \bibinfo {author} {\bibfnamefont {C.~R.}\
  \bibnamefont {Dean}},\ and\ \bibinfo {author} {\bibfnamefont
  {X.}~\bibnamefont {Roy}},\ }\bibfield  {title} {\bibinfo {title} {{Layered
  Antiferromagnetism Induces Large Negative Magnetoresistance in the van der
  Waals Semiconductor CrSBr}},\ }\href
  {https://doi.org/https://doi.org/10.1002/adma.202003240} {\bibfield
  {journal} {\bibinfo  {journal} {Advanced Materials}\ }\textbf {\bibinfo
  {volume} {32}},\ \bibinfo {pages} {2003240} (\bibinfo {year}
  {2020})}\BibitemShut {NoStop}%
\bibitem [{\citenamefont {Wilson}\ \emph {et~al.}(2021)\citenamefont {Wilson},
  \citenamefont {Lee}, \citenamefont {Cenker}, \citenamefont {Xie},
  \citenamefont {Dismukes}, \citenamefont {Telford}, \citenamefont {Fonseca},
  \citenamefont {Sivakumar}, \citenamefont {Dean}, \citenamefont {Cao},
  \citenamefont {Roy}, \citenamefont {Xu},\ and\ \citenamefont
  {Zhu}}]{Wilson:2021aa}%
  \BibitemOpen
  \bibfield  {author} {\bibinfo {author} {\bibfnamefont {N.~P.}\ \bibnamefont
  {Wilson}}, \bibinfo {author} {\bibfnamefont {K.}~\bibnamefont {Lee}},
  \bibinfo {author} {\bibfnamefont {J.}~\bibnamefont {Cenker}}, \bibinfo
  {author} {\bibfnamefont {K.}~\bibnamefont {Xie}}, \bibinfo {author}
  {\bibfnamefont {A.~H.}\ \bibnamefont {Dismukes}}, \bibinfo {author}
  {\bibfnamefont {E.~J.}\ \bibnamefont {Telford}}, \bibinfo {author}
  {\bibfnamefont {J.}~\bibnamefont {Fonseca}}, \bibinfo {author} {\bibfnamefont
  {S.}~\bibnamefont {Sivakumar}}, \bibinfo {author} {\bibfnamefont
  {C.}~\bibnamefont {Dean}}, \bibinfo {author} {\bibfnamefont {T.}~\bibnamefont
  {Cao}}, \bibinfo {author} {\bibfnamefont {X.}~\bibnamefont {Roy}}, \bibinfo
  {author} {\bibfnamefont {X.}~\bibnamefont {Xu}},\ and\ \bibinfo {author}
  {\bibfnamefont {X.}~\bibnamefont {Zhu}},\ }\bibfield  {title} {\bibinfo
  {title} {Interlayer electronic coupling on demand in a {2D} magnetic
  semiconductor},\ }\href {https://doi.org/10.1038/s41563-021-01070-8}
  {\bibfield  {journal} {\bibinfo  {journal} {Nature Materials}\ }\textbf
  {\bibinfo {volume} {20}},\ \bibinfo {pages} {1657} (\bibinfo {year}
  {2021})}\BibitemShut {NoStop}%
\bibitem [{\citenamefont {Klein}\ \emph {et~al.}(2023)\citenamefont {Klein},
  \citenamefont {Pingault}, \citenamefont {Florian}, \citenamefont
  {Hei{\ss}enb{\"u}ttel}, \citenamefont {Steinhoff}, \citenamefont {Song},
  \citenamefont {Torres}, \citenamefont {Dirnberger}, \citenamefont {Curtis},
  \citenamefont {Weile}, \citenamefont {Penn}, \citenamefont {Deilmann},
  \citenamefont {Dana}, \citenamefont {Bushati}, \citenamefont {Quan},
  \citenamefont {Luxa}, \citenamefont {Sofer}, \citenamefont {Al{\`u}},
  \citenamefont {Menon}, \citenamefont {Wurstbauer}, \citenamefont {Rohlfing},
  \citenamefont {Narang}, \citenamefont {Lon{\v c}ar},\ and\ \citenamefont
  {Ross}}]{Klein:2023aa}%
  \BibitemOpen
  \bibfield  {author} {\bibinfo {author} {\bibfnamefont {J.}~\bibnamefont
  {Klein}}, \bibinfo {author} {\bibfnamefont {B.}~\bibnamefont {Pingault}},
  \bibinfo {author} {\bibfnamefont {M.}~\bibnamefont {Florian}}, \bibinfo
  {author} {\bibfnamefont {M.-C.}\ \bibnamefont {Hei{\ss}enb{\"u}ttel}},
  \bibinfo {author} {\bibfnamefont {A.}~\bibnamefont {Steinhoff}}, \bibinfo
  {author} {\bibfnamefont {Z.}~\bibnamefont {Song}}, \bibinfo {author}
  {\bibfnamefont {K.}~\bibnamefont {Torres}}, \bibinfo {author} {\bibfnamefont
  {F.}~\bibnamefont {Dirnberger}}, \bibinfo {author} {\bibfnamefont {J.~B.}\
  \bibnamefont {Curtis}}, \bibinfo {author} {\bibfnamefont {M.}~\bibnamefont
  {Weile}}, \bibinfo {author} {\bibfnamefont {A.}~\bibnamefont {Penn}},
  \bibinfo {author} {\bibfnamefont {T.}~\bibnamefont {Deilmann}}, \bibinfo
  {author} {\bibfnamefont {R.}~\bibnamefont {Dana}}, \bibinfo {author}
  {\bibfnamefont {R.}~\bibnamefont {Bushati}}, \bibinfo {author} {\bibfnamefont
  {J.}~\bibnamefont {Quan}}, \bibinfo {author} {\bibfnamefont {J.}~\bibnamefont
  {Luxa}}, \bibinfo {author} {\bibfnamefont {Z.}~\bibnamefont {Sofer}},
  \bibinfo {author} {\bibfnamefont {A.}~\bibnamefont {Al{\`u}}}, \bibinfo
  {author} {\bibfnamefont {V.~M.}\ \bibnamefont {Menon}}, \bibinfo {author}
  {\bibfnamefont {U.}~\bibnamefont {Wurstbauer}}, \bibinfo {author}
  {\bibfnamefont {M.}~\bibnamefont {Rohlfing}}, \bibinfo {author}
  {\bibfnamefont {P.}~\bibnamefont {Narang}}, \bibinfo {author} {\bibfnamefont
  {M.}~\bibnamefont {Lon{\v c}ar}},\ and\ \bibinfo {author} {\bibfnamefont
  {F.~M.}\ \bibnamefont {Ross}},\ }\bibfield  {title} {\bibinfo {title} {{The
  Bulk van der Waals Layered Magnet {CrSBr} is a Quasi-1D Material}},\ }\href
  {https://doi.org/10.1021/acsnano.2c07316} {\bibfield  {journal} {\bibinfo
  {journal} {ACS Nano}\ }\textbf {\bibinfo {volume} {17}},\ \bibinfo {pages}
  {5316} (\bibinfo {year} {2023})}\BibitemShut {NoStop}%
\bibitem [{\citenamefont {Dirnberger}\ \emph {et~al.}(2023)\citenamefont
  {Dirnberger}, \citenamefont {Quan}, \citenamefont {Bushati}, \citenamefont
  {Diederich}, \citenamefont {Florian}, \citenamefont {Klein}, \citenamefont
  {Mosina}, \citenamefont {Sofer}, \citenamefont {Xu}, \citenamefont {Kamra},
  \citenamefont {Garc{\'\i}a-Vidal}, \citenamefont {Al{\`u}},\ and\
  \citenamefont {Menon}}]{Dirnberger:2023aa}%
  \BibitemOpen
  \bibfield  {author} {\bibinfo {author} {\bibfnamefont {F.}~\bibnamefont
  {Dirnberger}}, \bibinfo {author} {\bibfnamefont {J.}~\bibnamefont {Quan}},
  \bibinfo {author} {\bibfnamefont {R.}~\bibnamefont {Bushati}}, \bibinfo
  {author} {\bibfnamefont {G.~M.}\ \bibnamefont {Diederich}}, \bibinfo {author}
  {\bibfnamefont {M.}~\bibnamefont {Florian}}, \bibinfo {author} {\bibfnamefont
  {J.}~\bibnamefont {Klein}}, \bibinfo {author} {\bibfnamefont
  {K.}~\bibnamefont {Mosina}}, \bibinfo {author} {\bibfnamefont
  {Z.}~\bibnamefont {Sofer}}, \bibinfo {author} {\bibfnamefont
  {X.}~\bibnamefont {Xu}}, \bibinfo {author} {\bibfnamefont {A.}~\bibnamefont
  {Kamra}}, \bibinfo {author} {\bibfnamefont {F.~J.}\ \bibnamefont
  {Garc{\'\i}a-Vidal}}, \bibinfo {author} {\bibfnamefont {A.}~\bibnamefont
  {Al{\`u}}},\ and\ \bibinfo {author} {\bibfnamefont {V.~M.}\ \bibnamefont
  {Menon}},\ }\bibfield  {title} {\bibinfo {title} {Magneto-optics in a van der
  {Waals} magnet tuned by self-hybridized polaritons},\ }\href
  {https://doi.org/10.1038/s41586-023-06275-2} {\bibfield  {journal} {\bibinfo
  {journal} {Nature}\ }\textbf {\bibinfo {volume} {620}},\ \bibinfo {pages}
  {533} (\bibinfo {year} {2023})}\BibitemShut {NoStop}%
\bibitem [{\citenamefont {Klein}\ and\ \citenamefont {Ross}(2024)}]{Klein2024}%
  \BibitemOpen
  \bibfield  {author} {\bibinfo {author} {\bibfnamefont {J.}~\bibnamefont
  {Klein}}\ and\ \bibinfo {author} {\bibfnamefont {F.~M.}\ \bibnamefont
  {Ross}},\ }\bibfield  {title} {\bibinfo {title} {{Materials beyond
  monolayers: The magnetic quasi-1D semiconductor CrSBr}},\ }\bibfield
  {journal} {\bibinfo  {journal} {Journal of Materials Research}\ }\href
  {https://doi.org/10.1557/s43578-024-01459-6} {10.1557/s43578-024-01459-6}
  (\bibinfo {year} {2024})\BibitemShut {NoStop}%
\bibitem [{\citenamefont {Smolenski}\ \emph {et~al.}(2024)\citenamefont
  {Smolenski}, \citenamefont {Wen}, \citenamefont {Li}, \citenamefont {Downey},
  \citenamefont {Alfrey}, \citenamefont {Liu}, \citenamefont {Kondusamy},
  \citenamefont {Bostwick}, \citenamefont {Jozwiak}, \citenamefont {Rotenberg},
  \citenamefont {Zhao}, \citenamefont {Deng}, \citenamefont {Lv}, \citenamefont
  {Zgid}, \citenamefont {Gull},\ and\ \citenamefont
  {Jo}}]{smolenski2024largeexcitonbindingenergy}%
  \BibitemOpen
  \bibfield  {author} {\bibinfo {author} {\bibfnamefont {S.}~\bibnamefont
  {Smolenski}}, \bibinfo {author} {\bibfnamefont {M.}~\bibnamefont {Wen}},
  \bibinfo {author} {\bibfnamefont {Q.}~\bibnamefont {Li}}, \bibinfo {author}
  {\bibfnamefont {E.}~\bibnamefont {Downey}}, \bibinfo {author} {\bibfnamefont
  {A.}~\bibnamefont {Alfrey}}, \bibinfo {author} {\bibfnamefont
  {W.}~\bibnamefont {Liu}}, \bibinfo {author} {\bibfnamefont {A.~L.~N.}\
  \bibnamefont {Kondusamy}}, \bibinfo {author} {\bibfnamefont {A.}~\bibnamefont
  {Bostwick}}, \bibinfo {author} {\bibfnamefont {C.}~\bibnamefont {Jozwiak}},
  \bibinfo {author} {\bibfnamefont {E.}~\bibnamefont {Rotenberg}}, \bibinfo
  {author} {\bibfnamefont {L.}~\bibnamefont {Zhao}}, \bibinfo {author}
  {\bibfnamefont {H.}~\bibnamefont {Deng}}, \bibinfo {author} {\bibfnamefont
  {B.}~\bibnamefont {Lv}}, \bibinfo {author} {\bibfnamefont {D.}~\bibnamefont
  {Zgid}}, \bibinfo {author} {\bibfnamefont {E.}~\bibnamefont {Gull}},\ and\
  \bibinfo {author} {\bibfnamefont {N.~H.}\ \bibnamefont {Jo}},\ }\bibfield
  {title} {\bibinfo {title} {{Large Exciton Binding Energy in the Bulk van der
  Waals Magnet CrSBr}},\ }\href {https://arxiv.org/abs/2403.13897} {\bibfield
  {journal} {\bibinfo  {journal} {arXiv:2403.13897}\ } (\bibinfo {year}
  {2024})}\BibitemShut {NoStop}%
\bibitem [{\citenamefont {Watson}\ \emph {et~al.}(2024)\citenamefont {Watson},
  \citenamefont {Acharya}, \citenamefont {Nunn}, \citenamefont {Nagireddy},
  \citenamefont {Pashov}, \citenamefont {R{\"o}sner}, \citenamefont {van
  Schilfgaarde}, \citenamefont {Wilson},\ and\ \citenamefont
  {Cacho}}]{Watson2024}%
  \BibitemOpen
  \bibfield  {author} {\bibinfo {author} {\bibfnamefont {M.~D.}\ \bibnamefont
  {Watson}}, \bibinfo {author} {\bibfnamefont {S.}~\bibnamefont {Acharya}},
  \bibinfo {author} {\bibfnamefont {J.~E.}\ \bibnamefont {Nunn}}, \bibinfo
  {author} {\bibfnamefont {L.}~\bibnamefont {Nagireddy}}, \bibinfo {author}
  {\bibfnamefont {D.}~\bibnamefont {Pashov}}, \bibinfo {author} {\bibfnamefont
  {M.}~\bibnamefont {R{\"o}sner}}, \bibinfo {author} {\bibfnamefont
  {M.}~\bibnamefont {van Schilfgaarde}}, \bibinfo {author} {\bibfnamefont
  {N.~R.}\ \bibnamefont {Wilson}},\ and\ \bibinfo {author} {\bibfnamefont
  {C.}~\bibnamefont {Cacho}},\ }\bibfield  {title} {\bibinfo {title} {{Giant
  exchange splitting in the electronic structure of A-type 2D antiferromagnet
  CrSBr}},\ }\href {https://doi.org/10.1038/s41699-024-00492-7} {\bibfield
  {journal} {\bibinfo  {journal} {npj 2D Materials and Applications}\ }\textbf
  {\bibinfo {volume} {8}},\ \bibinfo {pages} {54} (\bibinfo {year}
  {2024})}\BibitemShut {NoStop}%
\bibitem [{\citenamefont {G{\"o}ser}\ \emph {et~al.}(1990)\citenamefont
  {G{\"o}ser}, \citenamefont {Paul},\ and\ \citenamefont
  {Kahle}}]{Goser:1990aa}%
  \BibitemOpen
  \bibfield  {author} {\bibinfo {author} {\bibfnamefont {O.}~\bibnamefont
  {G{\"o}ser}}, \bibinfo {author} {\bibfnamefont {W.}~\bibnamefont {Paul}},\
  and\ \bibinfo {author} {\bibfnamefont {H.~G.}\ \bibnamefont {Kahle}},\
  }\bibfield  {title} {\bibinfo {title} {Magnetic properties of {CrSBr}},\
  }\href {https://doi.org/https://doi.org/10.1016/0304-8853(90)90689-N}
  {\bibfield  {journal} {\bibinfo  {journal} {J. Magn. Magn. Mater.}\ }\textbf
  {\bibinfo {volume} {92}},\ \bibinfo {pages} {129} (\bibinfo {year}
  {1990})}\BibitemShut {NoStop}%
\bibitem [{\citenamefont {Scheie}\ \emph {et~al.}(2022)\citenamefont {Scheie},
  \citenamefont {Ziebel}, \citenamefont {Chica}, \citenamefont {Bae},
  \citenamefont {Wang}, \citenamefont {Kolesnikov}, \citenamefont {Zhu},\ and\
  \citenamefont {Roy}}]{Scheie:2022aa}%
  \BibitemOpen
  \bibfield  {author} {\bibinfo {author} {\bibfnamefont {A.}~\bibnamefont
  {Scheie}}, \bibinfo {author} {\bibfnamefont {M.}~\bibnamefont {Ziebel}},
  \bibinfo {author} {\bibfnamefont {D.~G.}\ \bibnamefont {Chica}}, \bibinfo
  {author} {\bibfnamefont {Y.~J.}\ \bibnamefont {Bae}}, \bibinfo {author}
  {\bibfnamefont {X.}~\bibnamefont {Wang}}, \bibinfo {author} {\bibfnamefont
  {A.~I.}\ \bibnamefont {Kolesnikov}}, \bibinfo {author} {\bibfnamefont
  {X.}~\bibnamefont {Zhu}},\ and\ \bibinfo {author} {\bibfnamefont
  {X.}~\bibnamefont {Roy}},\ }\bibfield  {title} {\bibinfo {title} {{Spin Waves
  and Magnetic Exchange Hamiltonian in CrSBr}},\ }\href
  {https://doi.org/https://doi.org/10.1002/advs.202202467} {\bibfield
  {journal} {\bibinfo  {journal} {Advanced Science}\ }\textbf {\bibinfo
  {volume} {9}},\ \bibinfo {pages} {2202467} (\bibinfo {year}
  {2022})}\BibitemShut {NoStop}%
\bibitem [{\citenamefont {Lee}\ \emph {et~al.}(2021)\citenamefont {Lee},
  \citenamefont {Dismukes}, \citenamefont {Telford}, \citenamefont {Wiscons},
  \citenamefont {Wang}, \citenamefont {Xu}, \citenamefont {Nuckolls},
  \citenamefont {Dean}, \citenamefont {Roy},\ and\ \citenamefont
  {Zhu}}]{Lee:2021aa}%
  \BibitemOpen
  \bibfield  {author} {\bibinfo {author} {\bibfnamefont {K.}~\bibnamefont
  {Lee}}, \bibinfo {author} {\bibfnamefont {A.~H.}\ \bibnamefont {Dismukes}},
  \bibinfo {author} {\bibfnamefont {E.~J.}\ \bibnamefont {Telford}}, \bibinfo
  {author} {\bibfnamefont {R.~A.}\ \bibnamefont {Wiscons}}, \bibinfo {author}
  {\bibfnamefont {J.}~\bibnamefont {Wang}}, \bibinfo {author} {\bibfnamefont
  {X.}~\bibnamefont {Xu}}, \bibinfo {author} {\bibfnamefont {C.}~\bibnamefont
  {Nuckolls}}, \bibinfo {author} {\bibfnamefont {C.~R.}\ \bibnamefont {Dean}},
  \bibinfo {author} {\bibfnamefont {X.}~\bibnamefont {Roy}},\ and\ \bibinfo
  {author} {\bibfnamefont {X.}~\bibnamefont {Zhu}},\ }\bibfield  {title}
  {\bibinfo {title} {{Magnetic Order and Symmetry in the 2D Semiconductor
  CrSBr}},\ }\href {https://doi.org/10.1021/acs.nanolett.1c00219} {\bibfield
  {journal} {\bibinfo  {journal} {Nano Letters}\ }\textbf {\bibinfo {volume}
  {21}},\ \bibinfo {pages} {3511} (\bibinfo {year} {2021})}\BibitemShut
  {NoStop}%
\bibitem [{\citenamefont {Bo}\ \emph {et~al.}(2023)\citenamefont {Bo},
  \citenamefont {Li}, \citenamefont {Xu}, \citenamefont {Wan},\ and\
  \citenamefont {Pu}}]{Bo_2023}%
  \BibitemOpen
  \bibfield  {author} {\bibinfo {author} {\bibfnamefont {X.}~\bibnamefont
  {Bo}}, \bibinfo {author} {\bibfnamefont {F.}~\bibnamefont {Li}}, \bibinfo
  {author} {\bibfnamefont {X.}~\bibnamefont {Xu}}, \bibinfo {author}
  {\bibfnamefont {X.}~\bibnamefont {Wan}},\ and\ \bibinfo {author}
  {\bibfnamefont {Y.}~\bibnamefont {Pu}},\ }\bibfield  {title} {\bibinfo
  {title} {{Calculated magnetic exchange interactions in the van der Waals
  layered magnet CrSBr}},\ }\href {https://doi.org/10.1088/1367-2630/acb3ee}
  {\bibfield  {journal} {\bibinfo  {journal} {New J. Phys.}\ }\textbf {\bibinfo
  {volume} {25}},\ \bibinfo {pages} {013026} (\bibinfo {year}
  {2023})}\BibitemShut {NoStop}%
\bibitem [{\citenamefont {Bae}\ \emph {et~al.}(2022)\citenamefont {Bae},
  \citenamefont {Wang}, \citenamefont {Scheie}, \citenamefont {Xu},
  \citenamefont {Chica}, \citenamefont {Diederich}, \citenamefont {Cenker},
  \citenamefont {Ziebel}, \citenamefont {Bai}, \citenamefont {Ren},
  \citenamefont {Dean}, \citenamefont {Delor}, \citenamefont {Xu},
  \citenamefont {Roy}, \citenamefont {Kent},\ and\ \citenamefont
  {Zhu}}]{Bae:2022aa}%
  \BibitemOpen
  \bibfield  {author} {\bibinfo {author} {\bibfnamefont {Y.~J.}\ \bibnamefont
  {Bae}}, \bibinfo {author} {\bibfnamefont {J.}~\bibnamefont {Wang}}, \bibinfo
  {author} {\bibfnamefont {A.}~\bibnamefont {Scheie}}, \bibinfo {author}
  {\bibfnamefont {J.}~\bibnamefont {Xu}}, \bibinfo {author} {\bibfnamefont
  {D.~G.}\ \bibnamefont {Chica}}, \bibinfo {author} {\bibfnamefont {G.~M.}\
  \bibnamefont {Diederich}}, \bibinfo {author} {\bibfnamefont {J.}~\bibnamefont
  {Cenker}}, \bibinfo {author} {\bibfnamefont {M.~E.}\ \bibnamefont {Ziebel}},
  \bibinfo {author} {\bibfnamefont {Y.}~\bibnamefont {Bai}}, \bibinfo {author}
  {\bibfnamefont {H.}~\bibnamefont {Ren}}, \bibinfo {author} {\bibfnamefont
  {C.~R.}\ \bibnamefont {Dean}}, \bibinfo {author} {\bibfnamefont
  {M.}~\bibnamefont {Delor}}, \bibinfo {author} {\bibfnamefont
  {X.}~\bibnamefont {Xu}}, \bibinfo {author} {\bibfnamefont {X.}~\bibnamefont
  {Roy}}, \bibinfo {author} {\bibfnamefont {A.~D.}\ \bibnamefont {Kent}},\ and\
  \bibinfo {author} {\bibfnamefont {X.}~\bibnamefont {Zhu}},\ }\bibfield
  {title} {\bibinfo {title} {Exciton-coupled coherent magnons in a {2D}
  semiconductor},\ }\href {https://doi.org/10.1038/s41586-022-05024-1}
  {\bibfield  {journal} {\bibinfo  {journal} {Nature}\ }\textbf {\bibinfo
  {volume} {609}},\ \bibinfo {pages} {282} (\bibinfo {year}
  {2022})}\BibitemShut {NoStop}%
\bibitem [{\citenamefont {Diederich}\ \emph {et~al.}(2023)\citenamefont
  {Diederich}, \citenamefont {Cenker}, \citenamefont {Ren}, \citenamefont
  {Fonseca}, \citenamefont {Chica}, \citenamefont {Bae}, \citenamefont {Zhu},
  \citenamefont {Roy}, \citenamefont {Cao}, \citenamefont {Xiao},\ and\
  \citenamefont {Xu}}]{Diederich:2023aa}%
  \BibitemOpen
  \bibfield  {author} {\bibinfo {author} {\bibfnamefont {G.~M.}\ \bibnamefont
  {Diederich}}, \bibinfo {author} {\bibfnamefont {J.}~\bibnamefont {Cenker}},
  \bibinfo {author} {\bibfnamefont {Y.}~\bibnamefont {Ren}}, \bibinfo {author}
  {\bibfnamefont {J.}~\bibnamefont {Fonseca}}, \bibinfo {author} {\bibfnamefont
  {D.~G.}\ \bibnamefont {Chica}}, \bibinfo {author} {\bibfnamefont {Y.~J.}\
  \bibnamefont {Bae}}, \bibinfo {author} {\bibfnamefont {X.}~\bibnamefont
  {Zhu}}, \bibinfo {author} {\bibfnamefont {X.}~\bibnamefont {Roy}}, \bibinfo
  {author} {\bibfnamefont {T.}~\bibnamefont {Cao}}, \bibinfo {author}
  {\bibfnamefont {D.}~\bibnamefont {Xiao}},\ and\ \bibinfo {author}
  {\bibfnamefont {X.}~\bibnamefont {Xu}},\ }\bibfield  {title} {\bibinfo
  {title} {Tunable interaction between excitons and hybridized magnons in a
  layered semiconductor},\ }\href {https://doi.org/10.1038/s41565-022-01259-1}
  {\bibfield  {journal} {\bibinfo  {journal} {Nature Nanotechnology}\ }\textbf
  {\bibinfo {volume} {18}},\ \bibinfo {pages} {23} (\bibinfo {year}
  {2023})}\BibitemShut {NoStop}%
\bibitem [{\citenamefont {Linhart}\ \emph {et~al.}(2023)\citenamefont
  {Linhart}, \citenamefont {Rybak}, \citenamefont {Birowska}, \citenamefont
  {Scharoch}, \citenamefont {Mosina}, \citenamefont {Mazanek}, \citenamefont
  {Kaczorowski}, \citenamefont {Sofer},\ and\ \citenamefont
  {Kudrawiec}}]{D3TC01216F}%
  \BibitemOpen
  \bibfield  {author} {\bibinfo {author} {\bibfnamefont {W.~M.}\ \bibnamefont
  {Linhart}}, \bibinfo {author} {\bibfnamefont {M.}~\bibnamefont {Rybak}},
  \bibinfo {author} {\bibfnamefont {M.}~\bibnamefont {Birowska}}, \bibinfo
  {author} {\bibfnamefont {P.}~\bibnamefont {Scharoch}}, \bibinfo {author}
  {\bibfnamefont {K.}~\bibnamefont {Mosina}}, \bibinfo {author} {\bibfnamefont
  {V.}~\bibnamefont {Mazanek}}, \bibinfo {author} {\bibfnamefont
  {D.}~\bibnamefont {Kaczorowski}}, \bibinfo {author} {\bibfnamefont
  {Z.}~\bibnamefont {Sofer}},\ and\ \bibinfo {author} {\bibfnamefont
  {R.}~\bibnamefont {Kudrawiec}},\ }\bibfield  {title} {\bibinfo {title}
  {Optical markers of magnetic phase transition in {CrSBr}},\ }\href
  {https://doi.org/10.1039/D3TC01216F} {\bibfield  {journal} {\bibinfo
  {journal} {J. Mater. Chem. C}\ }\textbf {\bibinfo {volume} {11}},\ \bibinfo
  {pages} {8423} (\bibinfo {year} {2023})}\BibitemShut {NoStop}%
\bibitem [{\citenamefont {Tabataba-Vakili}\ \emph {et~al.}(2024)\citenamefont
  {Tabataba-Vakili}, \citenamefont {Nguyen}, \citenamefont {Rupp},
  \citenamefont {Mosina}, \citenamefont {Papavasileiou}, \citenamefont
  {Watanabe}, \citenamefont {Taniguchi}, \citenamefont {Maletinsky},
  \citenamefont {Glazov}, \citenamefont {Sofer}, \citenamefont {Baimuratov},\
  and\ \citenamefont {H{\"o}gele}}]{Tabataba-Vakili:2024aa}%
  \BibitemOpen
  \bibfield  {author} {\bibinfo {author} {\bibfnamefont {F.}~\bibnamefont
  {Tabataba-Vakili}}, \bibinfo {author} {\bibfnamefont {H.~P.~G.}\ \bibnamefont
  {Nguyen}}, \bibinfo {author} {\bibfnamefont {A.}~\bibnamefont {Rupp}},
  \bibinfo {author} {\bibfnamefont {K.}~\bibnamefont {Mosina}}, \bibinfo
  {author} {\bibfnamefont {A.}~\bibnamefont {Papavasileiou}}, \bibinfo {author}
  {\bibfnamefont {K.}~\bibnamefont {Watanabe}}, \bibinfo {author}
  {\bibfnamefont {T.}~\bibnamefont {Taniguchi}}, \bibinfo {author}
  {\bibfnamefont {P.}~\bibnamefont {Maletinsky}}, \bibinfo {author}
  {\bibfnamefont {M.~M.}\ \bibnamefont {Glazov}}, \bibinfo {author}
  {\bibfnamefont {Z.}~\bibnamefont {Sofer}}, \bibinfo {author} {\bibfnamefont
  {A.~S.}\ \bibnamefont {Baimuratov}},\ and\ \bibinfo {author} {\bibfnamefont
  {A.}~\bibnamefont {H{\"o}gele}},\ }\bibfield  {title} {\bibinfo {title}
  {Doping-control of excitons and magnetism in few-layer {CrSBr}},\ }\href
  {https://doi.org/10.1038/s41467-024-49048-9} {\bibfield  {journal} {\bibinfo
  {journal} {Nature Communications}\ }\textbf {\bibinfo {volume} {15}},\
  \bibinfo {pages} {4735} (\bibinfo {year} {2024})}\BibitemShut {NoStop}%
\bibitem [{\citenamefont {Ivchenko}(2005)}]{ivchenko05a}%
  \BibitemOpen
  \bibfield  {author} {\bibinfo {author} {\bibfnamefont {E.~L.}\ \bibnamefont
  {Ivchenko}},\ }\href@noop {} {\emph {\bibinfo {title} {Optical spectroscopy
  of semiconductor nanostructures}}}\ (\bibinfo  {publisher} {Alpha Science,
  Harrow UK},\ \bibinfo {year} {2005})\BibitemShut {NoStop}%
\bibitem [{\citenamefont {Klingshirn}(2012)}]{Klingshirn2012}%
  \BibitemOpen
  \bibfield  {author} {\bibinfo {author} {\bibfnamefont {C.~F.}\ \bibnamefont
  {Klingshirn}},\ }\href {https://doi.org/10.1007/978-3-642-28362-8} {\emph
  {\bibinfo {title} {Semiconductor Optics}}}\ (\bibinfo  {publisher} {Springer
  Berlin Heidelberg},\ \bibinfo {year} {2012})\BibitemShut {NoStop}%
\bibitem [{\citenamefont {Yu}\ \emph {et~al.}(2015)\citenamefont {Yu},
  \citenamefont {Cui}, \citenamefont {Xu},\ and\ \citenamefont
  {Yao}}]{Yu30122014}%
  \BibitemOpen
  \bibfield  {author} {\bibinfo {author} {\bibfnamefont {H.}~\bibnamefont
  {Yu}}, \bibinfo {author} {\bibfnamefont {X.}~\bibnamefont {Cui}}, \bibinfo
  {author} {\bibfnamefont {X.}~\bibnamefont {Xu}},\ and\ \bibinfo {author}
  {\bibfnamefont {W.}~\bibnamefont {Yao}},\ }\bibfield  {title} {\bibinfo
  {title} {Valley excitons in two-dimensional semiconductors},\ }\href
  {https://doi.org/10.1093/nsr/nwu078} {\bibfield  {journal} {\bibinfo
  {journal} {National Science Review}\ }\textbf {\bibinfo {volume} {2}},\
  \bibinfo {pages} {57} (\bibinfo {year} {2015})}\BibitemShut {NoStop}%
\bibitem [{\citenamefont {Wang}\ \emph {et~al.}(2018)\citenamefont {Wang},
  \citenamefont {Chernikov}, \citenamefont {Glazov}, \citenamefont {Heinz},
  \citenamefont {Marie}, \citenamefont {Amand},\ and\ \citenamefont
  {Urbaszek}}]{RevModPhys.90.021001}%
  \BibitemOpen
  \bibfield  {author} {\bibinfo {author} {\bibfnamefont {G.}~\bibnamefont
  {Wang}}, \bibinfo {author} {\bibfnamefont {A.}~\bibnamefont {Chernikov}},
  \bibinfo {author} {\bibfnamefont {M.~M.}\ \bibnamefont {Glazov}}, \bibinfo
  {author} {\bibfnamefont {T.~F.}\ \bibnamefont {Heinz}}, \bibinfo {author}
  {\bibfnamefont {X.}~\bibnamefont {Marie}}, \bibinfo {author} {\bibfnamefont
  {T.}~\bibnamefont {Amand}},\ and\ \bibinfo {author} {\bibfnamefont
  {B.}~\bibnamefont {Urbaszek}},\ }\bibfield  {title} {\bibinfo {title}
  {Colloquium: Excitons in atomically thin transition metal dichalcogenides},\
  }\href {https://doi.org/10.1103/RevModPhys.90.021001} {\bibfield  {journal}
  {\bibinfo  {journal} {Rev. Mod. Phys.}\ }\textbf {\bibinfo {volume} {90}},\
  \bibinfo {pages} {021001} (\bibinfo {year} {2018})}\BibitemShut {NoStop}%
\bibitem [{\citenamefont {Durnev}\ and\ \citenamefont
  {Glazov}(2018)}]{Durnev_2018}%
  \BibitemOpen
  \bibfield  {author} {\bibinfo {author} {\bibfnamefont {M.~V.}\ \bibnamefont
  {Durnev}}\ and\ \bibinfo {author} {\bibfnamefont {M.~M.}\ \bibnamefont
  {Glazov}},\ }\bibfield  {title} {\bibinfo {title} {Excitons and trions in
  two-dimensional semiconductors based on transition metal dichalcogenides},\
  }\href {https://doi.org/10.3367/ufne.2017.07.038172} {\bibfield  {journal}
  {\bibinfo  {journal} {Physics-Uspekhi}\ }\textbf {\bibinfo {volume} {61}},\
  \bibinfo {pages} {825} (\bibinfo {year} {2018})}\BibitemShut {NoStop}%
\bibitem [{\citenamefont {Semina}\ and\ \citenamefont
  {Suris}(2022)}]{Semina_2022}%
  \BibitemOpen
  \bibfield  {author} {\bibinfo {author} {\bibfnamefont {M.~A.}\ \bibnamefont
  {Semina}}\ and\ \bibinfo {author} {\bibfnamefont {R.~A.}\ \bibnamefont
  {Suris}},\ }\bibfield  {title} {\bibinfo {title} {Localized excitons and
  trions in semiconductor nanosystems},\ }\href
  {https://doi.org/10.3367/ufne.2020.11.038867} {\bibfield  {journal} {\bibinfo
   {journal} {Physics-Uspekhi}\ }\textbf {\bibinfo {volume} {65}},\ \bibinfo
  {pages} {111} (\bibinfo {year} {2022})}\BibitemShut {NoStop}%
\bibitem [{\citenamefont {Bianchi}\ \emph {et~al.}(2023)\citenamefont
  {Bianchi}, \citenamefont {Acharya}, \citenamefont {Dirnberger}, \citenamefont
  {Klein}, \citenamefont {Pashov}, \citenamefont {Mosina}, \citenamefont
  {Sofer}, \citenamefont {Rudenko}, \citenamefont {Katsnelson}, \citenamefont
  {van Schilfgaarde}, \citenamefont {R\"osner},\ and\ \citenamefont
  {Hofmann}}]{PhysRevB.107.235107}%
  \BibitemOpen
  \bibfield  {author} {\bibinfo {author} {\bibfnamefont {M.}~\bibnamefont
  {Bianchi}}, \bibinfo {author} {\bibfnamefont {S.}~\bibnamefont {Acharya}},
  \bibinfo {author} {\bibfnamefont {F.}~\bibnamefont {Dirnberger}}, \bibinfo
  {author} {\bibfnamefont {J.}~\bibnamefont {Klein}}, \bibinfo {author}
  {\bibfnamefont {D.}~\bibnamefont {Pashov}}, \bibinfo {author} {\bibfnamefont
  {K.}~\bibnamefont {Mosina}}, \bibinfo {author} {\bibfnamefont
  {Z.}~\bibnamefont {Sofer}}, \bibinfo {author} {\bibfnamefont {A.~N.}\
  \bibnamefont {Rudenko}}, \bibinfo {author} {\bibfnamefont {M.~I.}\
  \bibnamefont {Katsnelson}}, \bibinfo {author} {\bibfnamefont
  {M.}~\bibnamefont {van Schilfgaarde}}, \bibinfo {author} {\bibfnamefont
  {M.}~\bibnamefont {R\"osner}},\ and\ \bibinfo {author} {\bibfnamefont
  {P.}~\bibnamefont {Hofmann}},\ }\bibfield  {title} {\bibinfo {title}
  {Paramagnetic electronic structure of crsbr: Comparison between ab initio
  {$GW$} theory and angle-resolved photoemission spectroscopy},\ }\href
  {https://doi.org/10.1103/PhysRevB.107.235107} {\bibfield  {journal} {\bibinfo
   {journal} {Phys. Rev. B}\ }\textbf {\bibinfo {volume} {107}},\ \bibinfo
  {pages} {235107} (\bibinfo {year} {2023})}\BibitemShut {NoStop}%
\bibitem [{\citenamefont {Suris}\ \emph {et~al.}(2001)\citenamefont {Suris},
  \citenamefont {Kochereshko}, \citenamefont {Astakhov}, \citenamefont
  {Yakovlev}, \citenamefont {Ossau}, \citenamefont {N{\"u}rnberger},
  \citenamefont {Faschinger}, \citenamefont {Landwehr}, \citenamefont
  {Wojtowicz}, \citenamefont {Karczewski},\ and\ \citenamefont
  {Kossut}}]{PSSB:PSSB343}%
  \BibitemOpen
  \bibfield  {author} {\bibinfo {author} {\bibfnamefont {R.}~\bibnamefont
  {Suris}}, \bibinfo {author} {\bibfnamefont {V.}~\bibnamefont {Kochereshko}},
  \bibinfo {author} {\bibfnamefont {G.}~\bibnamefont {Astakhov}}, \bibinfo
  {author} {\bibfnamefont {D.}~\bibnamefont {Yakovlev}}, \bibinfo {author}
  {\bibfnamefont {W.}~\bibnamefont {Ossau}}, \bibinfo {author} {\bibfnamefont
  {J.}~\bibnamefont {N{\"u}rnberger}}, \bibinfo {author} {\bibfnamefont
  {W.}~\bibnamefont {Faschinger}}, \bibinfo {author} {\bibfnamefont
  {G.}~\bibnamefont {Landwehr}}, \bibinfo {author} {\bibfnamefont
  {T.}~\bibnamefont {Wojtowicz}}, \bibinfo {author} {\bibfnamefont
  {G.}~\bibnamefont {Karczewski}},\ and\ \bibinfo {author} {\bibfnamefont
  {J.}~\bibnamefont {Kossut}},\ }\bibfield  {title} {\bibinfo {title} {Excitons
  and trions modified by interaction with a two-dimensional electron gas},\
  }\href
  {https://doi.org/10.1002/1521-3951(200110)227:2<343::AID-PSSB343>3.0.CO;2-W}
  {\bibfield  {journal} {\bibinfo  {journal} {physica status solidi (b)}\
  }\textbf {\bibinfo {volume} {227}},\ \bibinfo {pages} {343} (\bibinfo {year}
  {2001})}\BibitemShut {NoStop}%
\bibitem [{\citenamefont {Koudinov}\ \emph {et~al.}(2014)\citenamefont
  {Koudinov}, \citenamefont {Kehl}, \citenamefont {Rodina}, \citenamefont
  {Geurts}, \citenamefont {Wolverson},\ and\ \citenamefont
  {Karczewski}}]{PhysRevLett.112.147402}%
  \BibitemOpen
  \bibfield  {author} {\bibinfo {author} {\bibfnamefont {A.~V.}\ \bibnamefont
  {Koudinov}}, \bibinfo {author} {\bibfnamefont {C.}~\bibnamefont {Kehl}},
  \bibinfo {author} {\bibfnamefont {A.~V.}\ \bibnamefont {Rodina}}, \bibinfo
  {author} {\bibfnamefont {J.}~\bibnamefont {Geurts}}, \bibinfo {author}
  {\bibfnamefont {D.}~\bibnamefont {Wolverson}},\ and\ \bibinfo {author}
  {\bibfnamefont {G.}~\bibnamefont {Karczewski}},\ }\bibfield  {title}
  {\bibinfo {title} {Suris tetrons: Possible spectroscopic evidence for
  four-particle optical excitations of a two-dimensional electron gas},\ }\href
  {https://doi.org/10.1103/PhysRevLett.112.147402} {\bibfield  {journal}
  {\bibinfo  {journal} {Phys. Rev. Lett.}\ }\textbf {\bibinfo {volume} {112}},\
  \bibinfo {pages} {147402} (\bibinfo {year} {2014})}\BibitemShut {NoStop}%
\bibitem [{\citenamefont {Sidler}\ \emph {et~al.}(2016)\citenamefont {Sidler},
  \citenamefont {Back}, \citenamefont {Cotlet}, \citenamefont {Srivastava},
  \citenamefont {Fink}, \citenamefont {Kroner}, \citenamefont {Demler},\ and\
  \citenamefont {Imamoglu}}]{Sidler:2016aa}%
  \BibitemOpen
  \bibfield  {author} {\bibinfo {author} {\bibfnamefont {M.}~\bibnamefont
  {Sidler}}, \bibinfo {author} {\bibfnamefont {P.}~\bibnamefont {Back}},
  \bibinfo {author} {\bibfnamefont {O.}~\bibnamefont {Cotlet}}, \bibinfo
  {author} {\bibfnamefont {A.}~\bibnamefont {Srivastava}}, \bibinfo {author}
  {\bibfnamefont {T.}~\bibnamefont {Fink}}, \bibinfo {author} {\bibfnamefont
  {M.}~\bibnamefont {Kroner}}, \bibinfo {author} {\bibfnamefont
  {E.}~\bibnamefont {Demler}},\ and\ \bibinfo {author} {\bibfnamefont
  {A.}~\bibnamefont {Imamoglu}},\ }\bibfield  {title} {\bibinfo {title} {Fermi
  polaron-polaritons in charge-tunable atomically thin semiconductors},\ }\href
  {http://dx.doi.org/10.1038/nphys3949} {\bibfield  {journal} {\bibinfo
  {journal} {Nature Physics}\ }\textbf {\bibinfo {volume} {13}},\ \bibinfo
  {pages} {255} (\bibinfo {year} {2016})}\BibitemShut {NoStop}%
%
\bibitem{PhysRevB.93.115314}
{A.~Chaves, M.~Z. Mayers, F.~M. Peeters, and D.~R. Reichman, Theoretical investigation of electron-hole complexes in anisotropic
  two-dimensional materials, Phys. Rev. B {\bf 93}, 115314 (2016).}
%
\bibitem{PhysRevB.98.235401}
{M.~Van~der Donck and F.~M. Peeters, Excitonic complexes in anisotropic atomically thin two-dimensional
  materials: Black phosphorus and ${\mathrm{TiS}}_{3}$, Phys. Rev. B {\bf 98}, 235401 (2018).}
%
\bibitem [{\citenamefont {Rytova}(1967)}]{Rytova1967}%
  \BibitemOpen
  \bibfield  {author} {\bibinfo {author} {\bibfnamefont {N.~S.}\ \bibnamefont
  {Rytova}},\ }\bibfield  {title} {\bibinfo {title} {{Screened potential of a
  point charge in a thin film}},\ }\href@noop {} {\bibfield  {journal}
  {\bibinfo  {journal} {Proc. MSU, Phys., Astron.}\ }\textbf {\bibinfo {volume}
  {3}},\ \bibinfo {pages} {18} (\bibinfo {year} {1967})}\BibitemShut {NoStop}%
\bibitem [{\citenamefont {{Keldysh}}(1979)}]{1979JETPL..29..658K}%
  \BibitemOpen
  \bibfield  {author} {\bibinfo {author} {\bibfnamefont {L.~V.}\ \bibnamefont
  {{Keldysh}}},\ }\bibfield  {title} {\bibinfo {title} {{Coulomb interaction in
  thin semiconductor and semimetal films}},\ }\href@noop {} {\bibfield
  {journal} {\bibinfo  {journal} {JETP Lett.}\ }\textbf {\bibinfo {volume}
  {29}},\ \bibinfo {pages} {658} (\bibinfo {year} {1979})}\BibitemShut
  {NoStop}%
\bibitem [{\citenamefont {Cudazzo}\ \emph {et~al.}(2011)\citenamefont
  {Cudazzo}, \citenamefont {Tokatly},\ and\ \citenamefont
  {Rubio}}]{Cudazzo:2011a}%
  \BibitemOpen
  \bibfield  {author} {\bibinfo {author} {\bibfnamefont {P.}~\bibnamefont
  {Cudazzo}}, \bibinfo {author} {\bibfnamefont {I.~V.}\ \bibnamefont
  {Tokatly}},\ and\ \bibinfo {author} {\bibfnamefont {A.}~\bibnamefont
  {Rubio}},\ }\bibfield  {title} {\bibinfo {title} {Dielectric screening in
  two-dimensional insulators: Implications for excitonic and impurity states in
  graphane},\ }\href {https://doi.org/10.1103/PhysRevB.84.085406} {\bibfield
  {journal} {\bibinfo  {journal} {Phys. Rev. B}\ }\textbf {\bibinfo {volume}
  {84}},\ \bibinfo {pages} {085406} (\bibinfo {year} {2011})}\BibitemShut
  {NoStop}%
\bibitem [{\citenamefont {Berkelbach}\ \emph {et~al.}(2013)\citenamefont
  {Berkelbach}, \citenamefont {Hybertsen},\ and\ \citenamefont
  {Reichman}}]{PhysRevB.88.045318}%
  \BibitemOpen
  \bibfield  {author} {\bibinfo {author} {\bibfnamefont {T.~C.}\ \bibnamefont
  {Berkelbach}}, \bibinfo {author} {\bibfnamefont {M.~S.}\ \bibnamefont
  {Hybertsen}},\ and\ \bibinfo {author} {\bibfnamefont {D.~R.}\ \bibnamefont
  {Reichman}},\ }\bibfield  {title} {\bibinfo {title} {Theory of neutral and
  charged excitons in monolayer transition metal dichalcogenides},\ }\href
  {https://doi.org/10.1103/PhysRevB.88.045318} {\bibfield  {journal} {\bibinfo
  {journal} {Phys. Rev. B}\ }\textbf {\bibinfo {volume} {88}},\ \bibinfo
  {pages} {045318} (\bibinfo {year} {2013})}\BibitemShut {NoStop}%
\bibitem [{\citenamefont {Semina}(2019)}]{Semina:2019aa}%
  \BibitemOpen
  \bibfield  {author} {\bibinfo {author} {\bibfnamefont {M.~A.}\ \bibnamefont
  {Semina}},\ }\bibfield  {title} {\bibinfo {title} {Excitons and trions in
  bilayer van der {Waals} heterostructures},\ }\href
  {https://doi.org/10.1134/S1063783419110301} {\bibfield  {journal} {\bibinfo
  {journal} {Physics of the Solid State}\ }\textbf {\bibinfo {volume} {61}},\
  \bibinfo {pages} {2218} (\bibinfo {year} {2019})}\BibitemShut {NoStop}%
\bibitem [{\citenamefont {Asriyan}\ \emph {et~al.}(2019)\citenamefont
  {Asriyan}, \citenamefont {Kurbakov}, \citenamefont {Fedorov},\ and\
  \citenamefont {Lozovik}}]{PhysRevB.99.085108}%
  \BibitemOpen
  \bibfield  {author} {\bibinfo {author} {\bibfnamefont {N.~A.}\ \bibnamefont
  {Asriyan}}, \bibinfo {author} {\bibfnamefont {I.~L.}\ \bibnamefont
  {Kurbakov}}, \bibinfo {author} {\bibfnamefont {A.~K.}\ \bibnamefont
  {Fedorov}},\ and\ \bibinfo {author} {\bibfnamefont {Y.~E.}\ \bibnamefont
  {Lozovik}},\ }\bibfield  {title} {\bibinfo {title} {Optical probing in a
  bilayer dark-bright condensate system},\ }\href
  {https://doi.org/10.1103/PhysRevB.99.085108} {\bibfield  {journal} {\bibinfo
  {journal} {Phys. Rev. B}\ }\textbf {\bibinfo {volume} {99}},\ \bibinfo
  {pages} {085108} (\bibinfo {year} {2019})}\BibitemShut {NoStop}%
%
\bibitem{PhysRevB.92.245123}
{S.~Latini, T.~Olsen, and K.~S. Thygesen, Excitons in van der Waals heterostructures: The important role of dielectric screening, Phys. Rev. B {\bf 92}, 245123 (2015).}
\bibitem{2053-1583-4-2-022004} {K.~S. Thygesen, Calculating excitons, plasmons, and quasiparticles in {2D} materials and van der {Waals} heterostructures, 2D Materials {\bf 4}, 022004 (2017)}.
\bibitem{PhysRevB.98.125308}
{D.~Van~Tuan, M.~Yang, and H.~Dery, Coulomb interaction in monolayer transition-metal dichalcogenides, Phys. Rev. B {\bf 98}, 125308 (2018).}
\bibitem{Glazov:2018aa}
{M.~M. Glazov and A.~Chernikov, Breakdown of the static approximation for free carrier screening of   excitons in monolayer semiconductors, physica status solidi (b) {\bf 255}, 1800216 (2018).}
\bibitem{Scharf:2019aa} {B.~Scharf, D.~Van~Tuan, I.~{\v Z}uti{\'c}, and H.~Dery, Dynamical screening in monolayer transition-metal dichalcogenides and its manifestations in the exciton spectrum, Journal of Physics: Condensed Matter {\bf 31}, 203001 (2019).}
%
\bibitem [{\citenamefont {Courtade}\ \emph {et~al.}(2017)\citenamefont
  {Courtade}, \citenamefont {Semina}, \citenamefont {Manca}, \citenamefont
  {Glazov}, \citenamefont {Robert}, \citenamefont {Cadiz}, \citenamefont
  {Wang}, \citenamefont {Taniguchi}, \citenamefont {Watanabe}, \citenamefont
  {Pierre}, \citenamefont {Escoffier}, \citenamefont {Ivchenko}, \citenamefont
  {Renucci}, \citenamefont {Marie}, \citenamefont {Amand},\ and\ \citenamefont
  {Urbaszek}}]{Courtade:2017a}%
  \BibitemOpen
  \bibfield  {author} {\bibinfo {author} {\bibfnamefont {E.}~\bibnamefont
  {Courtade}}, \bibinfo {author} {\bibfnamefont {M.}~\bibnamefont {Semina}},
  \bibinfo {author} {\bibfnamefont {M.}~\bibnamefont {Manca}}, \bibinfo
  {author} {\bibfnamefont {M.~M.}\ \bibnamefont {Glazov}}, \bibinfo {author}
  {\bibfnamefont {C.}~\bibnamefont {Robert}}, \bibinfo {author} {\bibfnamefont
  {F.}~\bibnamefont {Cadiz}}, \bibinfo {author} {\bibfnamefont
  {G.}~\bibnamefont {Wang}}, \bibinfo {author} {\bibfnamefont {T.}~\bibnamefont
  {Taniguchi}}, \bibinfo {author} {\bibfnamefont {K.}~\bibnamefont {Watanabe}},
  \bibinfo {author} {\bibfnamefont {M.}~\bibnamefont {Pierre}}, \bibinfo
  {author} {\bibfnamefont {W.}~\bibnamefont {Escoffier}}, \bibinfo {author}
  {\bibfnamefont {E.~L.}\ \bibnamefont {Ivchenko}}, \bibinfo {author}
  {\bibfnamefont {P.}~\bibnamefont {Renucci}}, \bibinfo {author} {\bibfnamefont
  {X.}~\bibnamefont {Marie}}, \bibinfo {author} {\bibfnamefont
  {T.}~\bibnamefont {Amand}},\ and\ \bibinfo {author} {\bibfnamefont
  {B.}~\bibnamefont {Urbaszek}},\ }\bibfield  {title} {\bibinfo {title}
  {Charged excitons in monolayer {WSe}$_{2}$: Experiment and theory},\ }\href
  {https://doi.org/10.1103/PhysRevB.96.085302} {\bibfield  {journal} {\bibinfo
  {journal} {Phys. Rev. B}\ }\textbf {\bibinfo {volume} {96}},\ \bibinfo
  {pages} {085302} (\bibinfo {year} {2017})}\BibitemShut {NoStop}%
\bibitem [{\citenamefont {Lin}\ \emph {et~al.}(2022)\citenamefont {Lin},
  \citenamefont {Ziegler}, \citenamefont {Semina}, \citenamefont {Mamedov},
  \citenamefont {Watanabe}, \citenamefont {Taniguchi}, \citenamefont {Bange},
  \citenamefont {Chernikov}, \citenamefont {Glazov},\ and\ \citenamefont
  {Lupton}}]{Lin:2022aa}%
  \BibitemOpen
  \bibfield  {author} {\bibinfo {author} {\bibfnamefont {K.-Q.}\ \bibnamefont
  {Lin}}, \bibinfo {author} {\bibfnamefont {J.~D.}\ \bibnamefont {Ziegler}},
  \bibinfo {author} {\bibfnamefont {M.~A.}\ \bibnamefont {Semina}}, \bibinfo
  {author} {\bibfnamefont {J.~V.}\ \bibnamefont {Mamedov}}, \bibinfo {author}
  {\bibfnamefont {K.}~\bibnamefont {Watanabe}}, \bibinfo {author}
  {\bibfnamefont {T.}~\bibnamefont {Taniguchi}}, \bibinfo {author}
  {\bibfnamefont {S.}~\bibnamefont {Bange}}, \bibinfo {author} {\bibfnamefont
  {A.}~\bibnamefont {Chernikov}}, \bibinfo {author} {\bibfnamefont {M.~M.}\
  \bibnamefont {Glazov}},\ and\ \bibinfo {author} {\bibfnamefont {J.~M.}\
  \bibnamefont {Lupton}},\ }\bibfield  {title} {\bibinfo {title} {High-lying
  valley-polarized trions in {2D} semiconductors},\ }\href
  {https://doi.org/10.1038/s41467-022-33939-w} {\bibfield  {journal} {\bibinfo
  {journal} {Nature Communications}\ }\textbf {\bibinfo {volume} {13}},\
  \bibinfo {pages} {6980} (\bibinfo {year} {2022})}\BibitemShut {NoStop}%
\bibitem [{\citenamefont {Semina}\ \emph {et~al.}(2023)\citenamefont {Semina},
  \citenamefont {Mamedov},\ and\ \citenamefont
  {Glazov}}]{10.1093/oxfmat/itad004}%
  \BibitemOpen
  \bibfield  {author} {\bibinfo {author} {\bibfnamefont {M.~A.}\ \bibnamefont
  {Semina}}, \bibinfo {author} {\bibfnamefont {J.~V.}\ \bibnamefont
  {Mamedov}},\ and\ \bibinfo {author} {\bibfnamefont {M.~M.}\ \bibnamefont
  {Glazov}},\ }\bibfield  {title} {\bibinfo {title} {{Excitons and trions with
  negative effective masses in two-dimensional semiconductors}},\ }\href
  {https://doi.org/10.1093/oxfmat/itad004} {\bibfield  {journal} {\bibinfo
  {journal} {Oxford Open Materials Science}\ }\textbf {\bibinfo {volume} {3}},\
  \bibinfo {pages} {itad004} (\bibinfo {year} {2023})}\BibitemShut {NoStop}%
\bibitem [{\citenamefont {Sergeev}\ and\ \citenamefont
  {Suris}(2001)}]{Sergeev:2001aa}%
  \BibitemOpen
  \bibfield  {author} {\bibinfo {author} {\bibfnamefont {R.}~\bibnamefont
  {Sergeev}}\ and\ \bibinfo {author} {\bibfnamefont {R.}~\bibnamefont
  {Suris}},\ }\bibfield  {title} {\bibinfo {title} {Ground-state energy of
  {$X^-$ and $X^+$} trions in a two-dimensional quantum well at an arbitrary
  mass ratio},\ }\href {https://doi.org/10.1134/1.1366005} {\bibfield
  {journal} {\bibinfo  {journal} {Physics of the Solid State}\ }\textbf
  {\bibinfo {volume} {43}},\ \bibinfo {pages} {746} (\bibinfo {year}
  {2001})}\BibitemShut {NoStop}%
  %
  \bibitem{PhysRevB.71.201312}
{G.~V. Astakhov, D.~R. Yakovlev, V.~V. Rudenkov, P.~C.~M. Christianen,
  T.~Barrick, S.~A. Crooker, A.~B. Dzyubenko, W.~Ossau, J.~C. Maan,
  G.~Karczewski, and T.~Wojtowicz, Definitive observation of the dark triplet ground state of charged
  excitons in high magnetic fields, {Phys. Rev. B} {\bf 71}, 201312 (2005).}
  \bibitem{Semina:2008aa}
{M.~A. Semina, R.~A. Sergeev, and R.~A. Suris,
 The binding energy of excitons and {X$^+$} and {X$^-$} trions in
  one-dimensional systems, Semiconductors {\bf 42}, 1427 (2008).}
%
\bibitem [{\citenamefont {Li}\ \emph {et~al.}(2023)\citenamefont {Li},
  \citenamefont {Xie}, \citenamefont {Alfrey}, \citenamefont {Beach},
  \citenamefont {McLellan}, \citenamefont {Lu}, \citenamefont {Hu},
  \citenamefont {Liu}, \citenamefont {Dhale}, \citenamefont {Lv}, \citenamefont
  {Zhao}, \citenamefont {Sun},\ and\ \citenamefont
  {Deng}}]{li2023magneticexcitonpolaritonstronglycoupled}%
  \BibitemOpen
  \bibfield  {author} {\bibinfo {author} {\bibfnamefont {Q.}~\bibnamefont
  {Li}}, \bibinfo {author} {\bibfnamefont {X.}~\bibnamefont {Xie}}, \bibinfo
  {author} {\bibfnamefont {A.}~\bibnamefont {Alfrey}}, \bibinfo {author}
  {\bibfnamefont {C.~W.}\ \bibnamefont {Beach}}, \bibinfo {author}
  {\bibfnamefont {N.}~\bibnamefont {McLellan}}, \bibinfo {author}
  {\bibfnamefont {Y.}~\bibnamefont {Lu}}, \bibinfo {author} {\bibfnamefont
  {J.}~\bibnamefont {Hu}}, \bibinfo {author} {\bibfnamefont {W.}~\bibnamefont
  {Liu}}, \bibinfo {author} {\bibfnamefont {N.}~\bibnamefont {Dhale}}, \bibinfo
  {author} {\bibfnamefont {B.}~\bibnamefont {Lv}}, \bibinfo {author}
  {\bibfnamefont {L.}~\bibnamefont {Zhao}}, \bibinfo {author} {\bibfnamefont
  {K.}~\bibnamefont {Sun}},\ and\ \bibinfo {author} {\bibfnamefont
  {H.}~\bibnamefont {Deng}},\ }\bibfield  {title} {\bibinfo {title} {Magnetic
  exciton-polariton with strongly coupled atomic and photonic anisotropies},\
  }\href {https://arxiv.org/abs/2306.11265} {\bibfield  {journal} {\bibinfo
  {journal} {arXiv:2306.11265}\ } (\bibinfo {year} {2023})}\BibitemShut
  {NoStop}%
%
\bibitem{kumar2024trionsmonolayertransitionmetal}
{S.~S. Kumar, B.~C. Mulkerin, A.~Tiene, F.~M. Marchetti, M.~M. Parish, and
  J.~Levinsen,  Efficient calculation of trion energies in monolayer transition metal
  dichalcogenides, Phys. Rev. B {\bf 111}, 085404 (2025).}
%
\bibitem{PhysRevB.72.035332}
{A.~S. Bracker, E.~A. Stinaff, D.~Gammon, M.~E. Ware, J.~G. Tischler, D.~Park,
  D.~Gershoni, A.~V. Filinov, M.~Bonitz, F.~Peeters, and C.~Riva, Binding energies of positive and negative trions: From quantum wells to quantum dots, Phys. Rev. B {\bf 72}, 035332 (2005).}
%
\bibitem{PhysRevB.96.035131}
{M.~Van~der Donck, M.~Zarenia, and F.~M. Peeters,
  Excitons and trions in monolayer transition metal dichalcogenides: A
  comparative study between the multiband model and the quadratic single-band
  model, Phys. Rev. B {\bf 96}, 035131 (2017).}
%
\bibitem [{\citenamefont {Sergeev}\ and\ \citenamefont
  {Suris}(2003)}]{Sergeev2003}%
  \BibitemOpen
  \bibfield  {author} {\bibinfo {author} {\bibfnamefont {R.~A.}\ \bibnamefont
  {Sergeev}}\ and\ \bibinfo {author} {\bibfnamefont {R.~A.}\ \bibnamefont
  {Suris}},\ }\bibfield  {title} {\bibinfo {title} {The {$X^+$} trion in a
  system with spatial separation of the charge carriers},\ }\href
  {https://doi.org/10.1134/1.1619518} {\bibfield  {journal} {\bibinfo
  {journal} {Semiconductors}\ }\textbf {\bibinfo {volume} {37}},\ \bibinfo
  {pages} {1205} (\bibinfo {year} {2003})}\BibitemShut {NoStop}%
\bibitem [{\citenamefont {{Semina}}\ and\ \citenamefont
  {{Suris}}(2011)}]{2011JETPL..94..574S}%
  \BibitemOpen
  \bibfield  {author} {\bibinfo {author} {\bibfnamefont {M.~A.}\ \bibnamefont
  {{Semina}}}\ and\ \bibinfo {author} {\bibfnamefont {R.~A.}\ \bibnamefont
  {{Suris}}},\ }\bibfield  {title} {\bibinfo {title} {{Coulomb states in
  nanostructures, accidental degeneracy, and the Laplace-Runge-Lenz
  operator}},\ }\href {https://doi.org/10.1134/S0021364011190155} {\bibfield
  {journal} {\bibinfo  {journal} {JETP Letters}\ }\textbf {\bibinfo {volume}
  {94}},\ \bibinfo {pages} {574} (\bibinfo {year} {2011})}\BibitemShut
  {NoStop}%
\bibitem [{\citenamefont {Ossau}\ and\ \citenamefont
  {Suris}(2003)}]{suris:correlation}%
  \BibitemOpen
  \bibinfo {editor} {\bibfnamefont {W.}~\bibnamefont {Ossau}}\ and\ \bibinfo
  {editor} {\bibfnamefont {R.}~\bibnamefont {Suris}},\ eds.,\ \bibinfo {title}
  {Optical properties of 2d systems with interacting electrons}\ (\bibinfo
  {publisher} {NATO ASI},\ \bibinfo {year} {2003})\ Chap.\ \bibinfo {chapter}
  {R. A. Suris, {Correlation between trion and hole in Fermi distribution in
  process of trion photo-excitation in doped QWs}}\BibitemShut {NoStop}%
\bibitem [{\citenamefont {Efimkin}\ and\ \citenamefont
  {MacDonald}(2017)}]{PhysRevB.95.035417}%
  \BibitemOpen
  \bibfield  {author} {\bibinfo {author} {\bibfnamefont {D.~K.}\ \bibnamefont
  {Efimkin}}\ and\ \bibinfo {author} {\bibfnamefont {A.~H.}\ \bibnamefont
  {MacDonald}},\ }\bibfield  {title} {\bibinfo {title} {Many-body theory of
  trion absorption features in two-dimensional semiconductors},\ }\href
  {https://doi.org/10.1103/PhysRevB.95.035417} {\bibfield  {journal} {\bibinfo
  {journal} {Phys. Rev. B}\ }\textbf {\bibinfo {volume} {95}},\ \bibinfo
  {pages} {035417} (\bibinfo {year} {2017})}\BibitemShut {NoStop}%
\bibitem [{\citenamefont {Glazov}(2020)}]{Glazov:2020wf}%
  \BibitemOpen
  \bibfield  {author} {\bibinfo {author} {\bibfnamefont {M.~M.}\ \bibnamefont
  {Glazov}},\ }\bibfield  {title} {\bibinfo {title} {Optical properties of
  charged excitons in two-dimensional semiconductors},\ }\href
  {https://doi.org/10.1063/5.0012475} {\bibfield  {journal} {\bibinfo
  {journal} {J. Chem. Phys.}\ }\textbf {\bibinfo {volume} {153}},\ \bibinfo
  {pages} {034703} (\bibinfo {year} {2020})}\BibitemShut {NoStop}%
\bibitem [{\citenamefont {Iakovlev}\ and\ \citenamefont
  {Glazov}(2023)}]{Iakovlev_2023}%
  \BibitemOpen
  \bibfield  {author} {\bibinfo {author} {\bibfnamefont {Z.~A.}\ \bibnamefont
  {Iakovlev}}\ and\ \bibinfo {author} {\bibfnamefont {M.~M.}\ \bibnamefont
  {Glazov}},\ }\bibfield  {title} {\bibinfo {title} {Fermi polaron fine
  structure in strained van der {Waals} heterostructures},\ }\href
  {https://doi.org/10.1088/2053-1583/acdd81} {\bibfield  {journal} {\bibinfo
  {journal} {2D Materials}\ }\textbf {\bibinfo {volume} {10}},\ \bibinfo
  {pages} {035034} (\bibinfo {year} {2023})}\BibitemShut {NoStop}%
  \bibitem{PhysRevX.9.041019}
{O.~Cotle\ifmmode~\mbox{\c{t}}\else \c{t}\fi{}, F.~Pientka, R.~Schmidt,
  G.~Zarand, E.~Demler, and A.~Imamoglu, Transport of neutral optical excitations using electric fields, Phys. Rev. X {\bf 9}, 041019 (2019).}
\bibitem{PhysRevA.85.021602}
{R.~Schmidt, T.~Enss, V.~Pietil\"a, and E.~Demler, Fermi polarons in two dimensions.
\newblock {Phys. Rev. A} {\bf 85}, 021602 (2012).}
  \bibitem [{\citenamefont {My\ifmmode~\acute{s}\else \'{s}\fi{}liwy}\ and\
  \citenamefont {Jachymski}(2024)}]{PhysRevB.109.214208}%
  \BibitemOpen
  \bibfield  {author} {\bibinfo {author} {\bibfnamefont {K.}~\bibnamefont
  {My\ifmmode~\acute{s}\else \'{s}\fi{}liwy}}\ and\ \bibinfo {author}
  {\bibfnamefont {K.}~\bibnamefont {Jachymski}},\ }\bibfield  {title} {\bibinfo
  {title} {Long-range interacting {Fermi polaron}},\ }\href
  {https://doi.org/10.1103/PhysRevB.109.214208} {\bibfield  {journal} {\bibinfo
   {journal} {Phys. Rev. B}\ }\textbf {\bibinfo {volume} {109}},\ \bibinfo
  {pages} {214208} (\bibinfo {year} {2024})}\BibitemShut {NoStop}%
%
\bibitem{PhysRevB.108.125406}
{A.~Tiene, B.~C. Mulkerin, J.~Levinsen, M.~M. Parish, and F.~M. Marchetti, Crossover from exciton polarons to trions in doped two-dimensional   semiconductors at finite temperature, Phys. Rev. B {\bf 108}, 125406 (2023).}
\bibitem{PhysRevB.102.085304}
{F.~Rana, O.~Koksal, and C.~Manolatou, Many-body theory of the optical conductivity of excitons and trions  in two-dimensional materials, Phys. Rev. B {\bf 102}, 085304 (2020).}
\bibitem{PhysRevB.98.235203}
{Y.-C. Chang, S.-Y. Shiau, and M.~Combescot, Crossover from trion-hole complex to exciton-polaron in $n$-doped
  two-dimensional semiconductor quantum wells, Phys. Rev. B {\bf 98}, 235203 (2018).}
%
\bibitem [{\citenamefont {Landau}\ and\ \citenamefont
  {Lifshitz}(1977)}]{ll3_eng}%
  \BibitemOpen
  \bibfield  {author} {\bibinfo {author} {\bibfnamefont {L.~D.}\ \bibnamefont
  {Landau}}\ and\ \bibinfo {author} {\bibfnamefont {E.~M.}\ \bibnamefont
  {Lifshitz}},\ }\href@noop {} {\emph {\bibinfo {title} {Quantum Mechanics:
  Non-Relativistic Theory}}}\ (\bibinfo  {publisher} {Butterworth-Heinemann,
  Oxford},\ \bibinfo {year} {1977})\BibitemShut {NoStop}%
\bibitem [{\citenamefont {McGuire}(1966)}]{McGuire:1966aa}%
  \BibitemOpen
  \bibfield  {author} {\bibinfo {author} {\bibfnamefont {J.~B.}\ \bibnamefont
  {McGuire}},\ }\bibfield  {title} {\bibinfo {title} {{Interacting Fermions in
  One Dimension. II. Attractive Potential}},\ }\href
  {https://doi.org/10.1063/1.1704798} {\bibfield  {journal} {\bibinfo
  {journal} {J. Math. Phys.}\ }\textbf {\bibinfo {volume} {7}},\ \bibinfo
  {pages} {123} (\bibinfo {year} {1966})}\BibitemShut {NoStop}%
\bibitem [{\citenamefont {Klawunn}\ and\ \citenamefont
  {Recati}(2011)}]{PhysRevA.84.033607}%
  \BibitemOpen
  \bibfield  {author} {\bibinfo {author} {\bibfnamefont {M.}~\bibnamefont
  {Klawunn}}\ and\ \bibinfo {author} {\bibfnamefont {A.}~\bibnamefont
  {Recati}},\ }\bibfield  {title} {\bibinfo {title} {Fermi polaron in two
  dimensions: Importance of the two-body bound state},\ }\href
  {https://doi.org/10.1103/PhysRevA.84.033607} {\bibfield  {journal} {\bibinfo
  {journal} {Phys. Rev. A}\ }\textbf {\bibinfo {volume} {84}},\ \bibinfo
  {pages} {033607} (\bibinfo {year} {2011})}\BibitemShut {NoStop}%
\bibitem [{Note1()}]{Note1}%
  \BibitemOpen
  \bibinfo {note} {For significant $V_0\protect \mathcal D$ a simple
  expression~\protect \eqref {sigma:exciton} for the exciton self-energy may
  not be applicable due to strong interaction of the Fermi sea hole with the
  exciton.}\BibitemShut {Stop}%
\bibitem [{\citenamefont {Maialle}\ \emph {et~al.}(1993)\citenamefont
  {Maialle}, \citenamefont {de~Andrada~e Silva},\ and\ \citenamefont
  {Sham}}]{maialle93}%
  \BibitemOpen
  \bibfield  {author} {\bibinfo {author} {\bibfnamefont {M.}~\bibnamefont
  {Maialle}}, \bibinfo {author} {\bibfnamefont {E.}~\bibnamefont {de~Andrada~e
  Silva}},\ and\ \bibinfo {author} {\bibfnamefont {L.}~\bibnamefont {Sham}},\
  }\bibfield  {title} {\bibinfo {title} {Exciton spin dynamics in quantum
  wells},\ }\href {https://doi.org/10.1103/PhysRevB.47.15776} {\bibfield
  {journal} {\bibinfo  {journal} {Phys. Rev. B}\ }\textbf {\bibinfo {volume}
  {47}},\ \bibinfo {pages} {15776} (\bibinfo {year} {1993})}\BibitemShut
  {NoStop}%
\bibitem [{\citenamefont {Gupalov}\ \emph {et~al.}(1998)\citenamefont
  {Gupalov}, \citenamefont {Ivchenko},\ and\ \citenamefont
  {Kavokin}}]{goupalov98}%
  \BibitemOpen
  \bibfield  {author} {\bibinfo {author} {\bibfnamefont {S.~V.}\ \bibnamefont
  {Gupalov}}, \bibinfo {author} {\bibfnamefont {E.~L.}\ \bibnamefont
  {Ivchenko}},\ and\ \bibinfo {author} {\bibfnamefont {A.~V.}\ \bibnamefont
  {Kavokin}},\ }\bibfield  {title} {\bibinfo {title} {Fine structure of
  localized exciton levels in quantum wells},\ }\href@noop {} {\bibfield
  {journal} {\bibinfo  {journal} {JETP}\ }\textbf {\bibinfo {volume} {86}},\
  \bibinfo {pages} {388} (\bibinfo {year} {1998})}\BibitemShut {NoStop}%
\bibitem [{\citenamefont {Yu}\ \emph {et~al.}(2014)\citenamefont {Yu},
  \citenamefont {Liu}, \citenamefont {Gong}, \citenamefont {Xu},\ and\
  \citenamefont {Yao}}]{Yu:2014fk-1}%
  \BibitemOpen
  \bibfield  {author} {\bibinfo {author} {\bibfnamefont {H.}~\bibnamefont
  {Yu}}, \bibinfo {author} {\bibfnamefont {G.-B.}\ \bibnamefont {Liu}},
  \bibinfo {author} {\bibfnamefont {P.}~\bibnamefont {Gong}}, \bibinfo {author}
  {\bibfnamefont {X.}~\bibnamefont {Xu}},\ and\ \bibinfo {author}
  {\bibfnamefont {W.}~\bibnamefont {Yao}},\ }\bibfield  {title} {\bibinfo
  {title} {Dirac cones and {D}irac saddle points of bright excitons in
  monolayer transition metal dichalcogenides},\ }\href
  {http://dx.doi.org/10.1038/ncomms4876} {\bibfield  {journal} {\bibinfo
  {journal} {Nat Commun}\ }\textbf {\bibinfo {volume} {5}},\ \bibinfo {pages}
  {3876} (\bibinfo {year} {2014})}\BibitemShut {NoStop}%
\bibitem [{\citenamefont {Glazov}\ \emph {et~al.}(2014)\citenamefont {Glazov},
  \citenamefont {Amand}, \citenamefont {Marie}, \citenamefont {Lagarde},
  \citenamefont {Bouet},\ and\ \citenamefont {Urbaszek}}]{glazov2014exciton}%
  \BibitemOpen
  \bibfield  {author} {\bibinfo {author} {\bibfnamefont {M.~M.}\ \bibnamefont
  {Glazov}}, \bibinfo {author} {\bibfnamefont {T.}~\bibnamefont {Amand}},
  \bibinfo {author} {\bibfnamefont {X.}~\bibnamefont {Marie}}, \bibinfo
  {author} {\bibfnamefont {D.}~\bibnamefont {Lagarde}}, \bibinfo {author}
  {\bibfnamefont {L.}~\bibnamefont {Bouet}},\ and\ \bibinfo {author}
  {\bibfnamefont {B.}~\bibnamefont {Urbaszek}},\ }\bibfield  {title} {\bibinfo
  {title} {Exciton fine structure and spin decoherence in monolayers of
  transition metal dichalcogenides},\ }\href
  {https://doi.org/10.1103/physrevb.89.201302} {\bibfield  {journal} {\bibinfo
  {journal} {Phys. Rev. B}\ }\textbf {\bibinfo {volume} {89}},\ \bibinfo
  {pages} {201302} (\bibinfo {year} {2014})}\BibitemShut {NoStop}%
\bibitem [{\citenamefont {Yu}\ and\ \citenamefont
  {Wu}(2014)}]{PhysRevB.89.205303}%
  \BibitemOpen
  \bibfield  {author} {\bibinfo {author} {\bibfnamefont {T.}~\bibnamefont
  {Yu}}\ and\ \bibinfo {author} {\bibfnamefont {M.~W.}\ \bibnamefont {Wu}},\
  }\bibfield  {title} {\bibinfo {title} {Valley depolarization due to
  intervalley and intravalley electron-hole exchange interactions in monolayer
  $\mbox{MoS}_{2}$},\ }\href {https://doi.org/10.1103/PhysRevB.89.205303}
  {\bibfield  {journal} {\bibinfo  {journal} {Phys. Rev. B}\ }\textbf {\bibinfo
  {volume} {89}},\ \bibinfo {pages} {205303} (\bibinfo {year}
  {2014})}\BibitemShut {NoStop}%
\bibitem [{\citenamefont {Fang}\ \emph {et~al.}(2019)\citenamefont {Fang},
  \citenamefont {Han}, \citenamefont {Robert}, \citenamefont {Semina},
  \citenamefont {Lagarde}, \citenamefont {Courtade}, \citenamefont {Taniguchi},
  \citenamefont {Watanabe}, \citenamefont {Amand}, \citenamefont {Urbaszek},
  \citenamefont {Glazov},\ and\ \citenamefont
  {Marie}}]{PhysRevLett.123.067401}%
  \BibitemOpen
  \bibfield  {author} {\bibinfo {author} {\bibfnamefont {H.~H.}\ \bibnamefont
  {Fang}}, \bibinfo {author} {\bibfnamefont {B.}~\bibnamefont {Han}}, \bibinfo
  {author} {\bibfnamefont {C.}~\bibnamefont {Robert}}, \bibinfo {author}
  {\bibfnamefont {M.~A.}\ \bibnamefont {Semina}}, \bibinfo {author}
  {\bibfnamefont {D.}~\bibnamefont {Lagarde}}, \bibinfo {author} {\bibfnamefont
  {E.}~\bibnamefont {Courtade}}, \bibinfo {author} {\bibfnamefont
  {T.}~\bibnamefont {Taniguchi}}, \bibinfo {author} {\bibfnamefont
  {K.}~\bibnamefont {Watanabe}}, \bibinfo {author} {\bibfnamefont
  {T.}~\bibnamefont {Amand}}, \bibinfo {author} {\bibfnamefont
  {B.}~\bibnamefont {Urbaszek}}, \bibinfo {author} {\bibfnamefont {M.~M.}\
  \bibnamefont {Glazov}},\ and\ \bibinfo {author} {\bibfnamefont
  {X.}~\bibnamefont {Marie}},\ }\bibfield  {title} {\bibinfo {title} {{Control
  of the Exciton Radiative Lifetime in van der Waals Heterostructures}},\
  }\href {https://doi.org/10.1103/PhysRevLett.123.067401} {\bibfield  {journal}
  {\bibinfo  {journal} {Phys. Rev. Lett.}\ }\textbf {\bibinfo {volume} {123}},\
  \bibinfo {pages} {067401} (\bibinfo {year} {2019})}\BibitemShut {NoStop}%
\bibitem [{\citenamefont {Iakovlev}\ and\ \citenamefont
  {Glazov}(2024)}]{Iakovlev:2024aa}%
  \BibitemOpen
  \bibfield  {author} {\bibinfo {author} {\bibfnamefont {Z.~A.}\ \bibnamefont
  {Iakovlev}}\ and\ \bibinfo {author} {\bibfnamefont {M.~M.}\ \bibnamefont
  {Glazov}},\ }\bibfield  {title} {\bibinfo {title} {Longitudinal-transverse
  splitting and fine structure of {Fermi} polarons in two-dimensional
  semiconductors},\ }\href
  {https://doi.org/https://doi.org/10.1016/j.jlumin.2024.120700} {\bibfield
  {journal} {\bibinfo  {journal} {J. Lumin.}\ }\textbf {\bibinfo {volume}
  {273}},\ \bibinfo {pages} {120700} (\bibinfo {year} {2024})}\BibitemShut
  {NoStop}%
\bibitem [{\citenamefont {Cohen}\ \emph {et~al.}(2024)\citenamefont {Cohen},
  \citenamefont {Haber}, \citenamefont {Neaton}, \citenamefont {Qiu},\ and\
  \citenamefont {Refaely-Abramson}}]{PhysRevLett.132.126902}%
  \BibitemOpen
  \bibfield  {author} {\bibinfo {author} {\bibfnamefont {G.}~\bibnamefont
  {Cohen}}, \bibinfo {author} {\bibfnamefont {J.~B.}\ \bibnamefont {Haber}},
  \bibinfo {author} {\bibfnamefont {J.~B.}\ \bibnamefont {Neaton}}, \bibinfo
  {author} {\bibfnamefont {D.~Y.}\ \bibnamefont {Qiu}},\ and\ \bibinfo {author}
  {\bibfnamefont {S.}~\bibnamefont {Refaely-Abramson}},\ }\bibfield  {title}
  {\bibinfo {title} {Phonon-driven femtosecond dynamics of excitons in
  crystalline pentacene from first principles},\ }\href
  {https://doi.org/10.1103/PhysRevLett.132.126902} {\bibfield  {journal}
  {\bibinfo  {journal} {Phys. Rev. Lett.}\ }\textbf {\bibinfo {volume} {132}},\
  \bibinfo {pages} {126902} (\bibinfo {year} {2024})}\BibitemShut {NoStop}%
\bibitem [{\citenamefont {Glazov}\ and\ \citenamefont
  {Suris}(2024)}]{glazov2024ultrafastexcitontransportvan}%
  \BibitemOpen
  \bibfield  {author} {\bibinfo {author} {\bibfnamefont {M.~M.}\ \bibnamefont
  {Glazov}}\ and\ \bibinfo {author} {\bibfnamefont {R.~A.}\ \bibnamefont
  {Suris}},\ }\bibfield  {title} {\bibinfo {title} {Ultrafast exciton transport
  in van der Waals heterostructures},\ }\href
  {https://arxiv.org/abs/2403.19571} {\bibfield  {journal} {\bibinfo  {journal}
  {Zh. Exp. Teor. Fiz.}\ }\textbf {\bibinfo {volume} {166}},\ \bibinfo {pages}
  {20} (\bibinfo {year} {2024})}\BibitemShut {NoStop}%
\bibitem [{\citenamefont {Loudon}(1959)}]{loudon59}%
  \BibitemOpen
  \bibfield  {author} {\bibinfo {author} {\bibfnamefont {R.}~\bibnamefont
  {Loudon}},\ }\bibfield  {title} {\bibinfo {title} {{One-Dimensional Hydrogen
  Atom}},\ }\href {https://doi.org/10.1119/1.1934950} {\bibfield  {journal}
  {\bibinfo  {journal} {American Journal of Physics}\ }\textbf {\bibinfo
  {volume} {27}},\ \bibinfo {pages} {649} (\bibinfo {year} {1959})}\BibitemShut
  {NoStop}%
\end{thebibliography}
%

\end{document}